\documentclass[a4paper,fleqn,usenatbib]{mnras}

\usepackage[T1]{fontenc}
\usepackage{ae,aecompl}

\usepackage{graphicx}	
\usepackage{amsmath}	
\usepackage{amssymb}	

\newcommand\rs[1]{_\mathrm{#1}}
\newcommand\avg[1]{\left<#1\right>}
\newcommand\opGm{\mathop{\Gamma}}
\newcommand\pd{\partial}
\newcommand\Prob{{\cal P}}
\newcommand\hateq{\mathrel{\hat=}}
\newcommand\Ord{{\cal O}}

\newcommand\al{\alpha}
\newcommand\bt{\beta}
\newcommand\gm{\gamma}
\newcommand\dl{\delta}
\newcommand\tht{\theta}
\newcommand\lmb{\lambda}
\newcommand\sg{\sigma}
\newcommand\om{\omega}
\newcommand\Gm{\Gamma}
\newcommand\Tht{\Theta}
\newcommand\phiz{\phi\rs{o}}

\newcommand\sintht{\sin\tht}
\newcommand\costht{\cos\tht}
\newcommand\sinsqtht{\sin^2\tht}
\newcommand\cossqtht{\cos^2\tht}
\newcommand\sinphi{\sin\varphi}
\newcommand\cosphi{\cos\varphi}
\newcommand\sinTht{\sin\Tht}
\newcommand\cosTht{\cos\Tht}
\newcommand\sinphiz{\sin\phiz}
\newcommand\cosphiz{\cos\phiz}
\newcommand\cosal{\cos\al}
\newcommand\sintau{\sin\tau}

\newcommand\me{m\rs{e}}

\newcommand\rad{r}
\newcommand\radproj{\varrho}
\newcommand\tradproj{\tilde\radproj}

\newcommand\rhoamb{\rho\rs{amb}}
\newcommand\Bamb{B\rs{amb}}
\newcommand\nz{n\rs{o}}
\newcommand\Bz{B\rs{o}}

\newcommand\Rs{R\rs{s}}
\newcommand\rhos{\rho\rs{s}}
\newcommand\comprratio{\kappa}

\newcommand\Wz{W\rs{o}}
\newcommand\As{A\rs{s}}

\newcommand\sgeff{\sg\rs{eff}}
\newcommand\fanis{f\rs{an}}
\newcommand\omc{\om\rs{c}}
\newcommand\omcsg{\om\rs{c,\sg}}

\newcommand\zobs{z\rs{obs}}

\newcommand\thtntwo{\tht_{Bn2}}
\newcommand\sinthtntwo{\sin\thtntwo}
\newcommand\costhtntwo{\cos\thtntwo}
\newcommand\sinsqthtntwo{\sin^2\thtntwo}
\newcommand\cossqthtntwo{\cos^2\thtntwo}
\newcommand\thtnone{\tht_{Bn1}}

\newcommand\Polperp{P_\perp}
\newcommand\Polpar{P_\parallel}
\newcommand\Pimax{\Pi\rs{max}}

\newcommand\Bav{\bar B}
\newcommand\Bperp{B_\perp}
\newcommand\Bpar{B_z}
\newcommand\vectB{\mathbfit{B}}
\newcommand\vectBav{\bar{\mathbfit{B}}}
\newcommand\vectdB{\dl\mathbfit{B}}
\newcommand\unitBav{\hat{\mathbfit{B}}}

\newcommand\Iloc{{\cal I}}
\newcommand\Qloc{{\cal Q}}
\newcommand\Uloc{{\cal U}}
\newcommand\Vloc{{\cal V}}
\newcommand\IlocPL{{\cal I}\rs{PL}}
\newcommand\QlocPL{{\cal Q}\rs{PL}}
\newcommand\QlocsqPL{{\cal Q}^2\rs{PL}}
\newcommand\UlocsqPL{{\cal U}^2\rs{PL}}
\newcommand\IlocfPL{{\cal I}'\rs{PL}}
\newcommand\QlocfPL{{\cal Q}'\rs{PL}}
\newcommand\avgIlocPL{\avg{{\cal I}}\rs{PL}}
\newcommand\avgQlocPL{\avg{{\cal Q}}\rs{PL}}
\newcommand\avgIlocPLasymp{\avg{{\cal I}}\rs{PL,asymp}}
\newcommand\avgQlocPLasymp{\avg{{\cal Q}}\rs{PL,asymp}}

\newcommand\Qi{Q\rs{i}}
\newcommand\Ui{U\rs{i}}
\newcommand\Qin[1]{Q_{\mathrm{i},#1}}
\newcommand\Uin[1]{U_{\mathrm{i},#1}}
\newcommand\Qf{Q\rs{f}}
\newcommand\Uf{U\rs{f}}
\newcommand\Qr{Q\rs{r}}
\newcommand\Ur{U\rs{r}}
\newcommand\RM{R\kern-.08em M}
\newcommand\RMf{R\kern-.08em M\rs{f}}
\newcommand\RMr{R\kern-.08em M\rs{r}}
\newcommand\RMx{R\kern-.08em M\rs{x}}
\newcommand\RMt{R\kern-.08em M\rs{t}}
\newcommand\QUm{Q\kern-.06em U\rs{m}}
\newcommand\QUp{Q\kern-.06em U\rs{p}}
\newcommand\fobs{f\rs{obs}}
\newcommand\fobstwo{f\rs{obs,2}}
\newcommand\RMobs{R\kern-.08em M\rs{obs}}
\newcommand\RMobszero{R\kern-.08em M\rs{obs,0}}
\newcommand\RMobstwo{R\kern-.08em M\rs{obs,2}}

\title[Radio polarization maps of SNRs]{Radio polarization maps of shell-type SNRs\\
{\LARGE I. Effects of a random magnetic field component, and thin-shell models}}

\author[R. Bandiera, O. Petruk]{
R. Bandiera,$^{1}$\thanks{E-mail: bandiera@arcetri.astro.it}
O. Petruk$^{2}$
\\
$^{1}$INAF - Osservatorio Astrofisico di Arcetri, Largo E. Fermi 5, 50125 Firenze, Italy\\
$^{2}$Institute for Applied Problems in Mechanics and Mathematics, Naukova Street, 3-b Lviv 79060, Ukraine
}

\date{Accepted 2016 March 3. Received 2016 February 28; in original form 2015 December 15}

\pubyear{2016}

\begin{document}
\label{firstpage}
\pagerange{\pageref{firstpage}--\pageref{lastpage}}
\maketitle


\begin{abstract}
The maps of intensity and polarization of the radio synchrotron emission from shell-type supernova remnants (SNRs) contain a considerable amount of information, although of not easy interpretation.
With the aim of deriving constraints on the 3-D spatial distribution of the emissivity, as well as on the structure of both ordered and random magnetic fields (MFs), we present here a scheme to model maps of the emission and polarization in SNRs.

We first generalize the classical treatment of the synchrotron emission to the case in which the MF is composed by an ordered MF plus an isotropic random component, with arbitrary relative strengths.
In the case of a power-law particle energy distribution, we derive analytic formulae that formally resemble those for the classical case.
We also treat the case of a shock compression of a fully random upstream field and we predict that the polarization fraction in this case should be higher than typically measured in SNRs.
We implement the above treatment into a code, which simulates the observed polarized emission of an emitting shell, taking into account also the effect of the internal Faraday rotation.

Finally, we show simulated maps for different orientations with respect to the observer, levels of the turbulent MF component, Faraday rotation levels, distributions of the emissivity (either barrel-shaped or limited to polar caps), and geometries for the ordered MF component (either tangential to the shell, or radial).
Their analysis allows us to outline properties useful for the interpretation of radio intensity and polarization maps.
\end{abstract}



\section{Introduction}

Supernova remnants (SNRs), an aftermath of stellar explosions, are among the first sources that have been observed by radio astronomers and, since the advent of radio interferometers, many of them have been mapped with high spatial resolution: their typical shell-like emission pattern roughly traces the location where the blast wave of the explosion hits the ambient medium.
In many cases radio detectors have been devised to provide the observer with full information about the incoming polarized radiation (i.e.~all four Stokes parameters).

In spite of this wealth of information, it seems that radio observations of SNRs have not been fully exploited, at least if compared with observations in other spectral bands, like for instance in the X rays.
One possible reason is that the radio emissivity, which in SNRs is typically non-thermal and of synchrotron origin, depends on a combination of physical parameters that are separately poorly known, namely the injection efficiency of the relativistic electrons and the magnitude of the magnetic fields (hereafter MFs).

On the other hand, the spatial structure of the polarization may provide important clues on the geometrical structure of the MF. For instance, it was noted \citep[][]{Milne-1987} that there is the tendency, in younger SNRs, for the polarization to be consistent with a predominantly radial structure of the MF while, in older SNRs, with a MF structure predominantly tangential to the SNR boundary: a review on this issue may be found in \citet{Dubner-Giacani-2015}.
This kind of dichotomy had been interpreted by some authors \citep[e.g.][]{Jun-Norman-1996, Inoue-etal-2013} as an evidence that in younger SNRs instabilities lead to strongly tangled MFs.
Instead in older SNRs the ordered component of the MF still keeps memory of the pattern originated by the shock compression of the ambient MF.
It has also been shown \citep[][]{Petruk-etal-2016} that in post-adiabatic SNRs the parallel component of the MF decreases, while the perpendicular one increases in the shock downstream.
Therefore, shocks of different obliquities in old SNRs tend to eventually become quasi-perpendicular, even those that initially were quasi-parallel.

MF turbulent amplification generated by kinetic processes in the case of a ``quasi-parallel'' shock \citep[e.g.][and reference therein]{Caprioli-Spitkovsky-2014b} should give rise to a tangential ordered component: in fact, efficient MF amplification taking place upstream of the shock leads, in the most extreme case, to a completely random MF; but then the shock compresses this field, enhancing the tangential component with respect to the radial one.
Therefore the observation of a radially oriented MFs cannot be justified even in this case, but suggests instead the onset of hydrodynamic instabilities, like Rayleigh-Taylor \citep[e.g.][]{Jun-Norman-1996} or Richtmyer-Meshkov \citep[e.g.][]{Inoue-etal-2013} instabilities.

Observations show that the polarization fractions vary from object to object, and from region to region within the same object, but their typical values are much lower than the maximum theoretically allowed limit ($\simeq69\%$, for a slope --0.5 of the synchrotron emission, namely a slope --2.0 in the energy distribution of the emitting electrons).
For instance, \citet{Dickel-Milne-1976} quote generally low polarizations, usually $<10\%$; while \citet{Kothes-etal-2006} show that the peaks of polarization fraction typically range from about $10\%$ to about $50\%$.
This seems to indicate that the discrepancy between theory and observations can be partly explained with the superposition, along the line of sight, of regions where the MFs have different orientations, and/or with the presence of a partially random MF.

This work presents a detailed treatment of synchrotron emission in the case of a partially random MF, and implements it within a thin-shell SNR model.
This approximation allows us to considerably simplify the problem (also making our numerical models much lighter to compute), while retaining a a large number of effects, and therefore allowing an analysis a wide variety of realistic cases.
Since our aim is to  to show in which way quantitative information could be effectively extracted from (suitably detailed) polarization maps, but not to model specific sources, we will discuss here only simple geometries.

The plan of the paper is as follows.

Sect.~2 analyzes the synchrotron emissivity and its polarization properties.
We begin reviewing the classical approach, valid for the case of particles with a power-law energy distribution (as it is typically the case for radio emitting particles) sitting in a uniform MF.
We then introduce a generalization of this approach, which extends its validity to the case of a partially disordered MF, with the random component assumed to be isotropic.
This treatment allows a continuity from the case of an almost uniform MF to that of an almost completely randomized one.
In a similar way, we also estimate the level of polarization induced by the compression, at a shock front, of an originally fully random MF.
Just a feeling is given of the difficulties that arise if the particle distribution is not a power-law, by considering the mono-energetic case.

For a correct modelling of radio polarization maps, to be compared with actual observations, it is required to properly sum up the local emissivity in the various Stokes parameters, by also correcting for the Faraday Rotation (FR) effects.
While this could apply to many different classes of astrophysical sources, in Sect.~3 we focus on the case of radio shell-type SNRs, with suitable assumptions on the geometry of the MF as well as on the spatial distribution of thermal gas and of the emitting particles.
In particular, we assume axial symmetry, and the ``thin-layer'' simplified approach, deriving formulae that we then extensively use in our modelling.
At the end of this section we also present a momenta approach, which allows us to formally treat also a more general case of source structure, in the limit of sufficiently short radiation wavelengths.

In Sect.~4 we present a variety of simulated maps obtained under different parameters choices, and we use them to discuss for instance the effects of changing the aspect angle, the level of internal FR, the level of MF fluctuations.

In addition to the ``classical'' case, in which the ordered MF component has a meridional structure, consistent with a compression of a pre-existing uniform ambient MF, in Sect.~5 we also investigate the completely different case, probably more appropriate for some young SNRs, in which the observed MF structure is mostly radial.
We compare some of your results with the observed structure of the SNR SN~1006.

Sect.~5 concludes.

\section{Local emissivity and polarization}
\label{polariz:synch-emiss-local}

In this section we study the synchrotron emissivity, from an element of volume.
The meaning of adding the word ``radio'' is that we consider only the case in which the energy distribution of the relevant electrons is a pure power law, as it is typically for radio emission.
For X-ray emitting electrons, instead, their energy distribution experiences a cutoff, which would require a different treatment from that presented below.

\subsection{The classical case of a uniform MF}
\label{polariz:emiss-orderedMF}

Let us begin with a review of the basic formulae about synchrotron radiation, and polarized radiation in general (see e.g. Rybicki \& Lightman 1979).
Let us first consider the case of a particle with a given Lorentz factor $\gm$, moving in a MF $\vectB$ (assumed to be uniform, at least on scales smaller than the particle gyration radius).
In this case the synchrotron power emitted per unit frequency ($\om=2\pi\nu$; so that $P(\nu)=2\pi P(\om)$) is the sum of polarized components, respectively perpendicular and parallel to the direction of the projected MF, which are equal to
\begin{eqnarray}
\Polperp(\om)&=&\frac{\sqrt{3}\,e^3\Bperp}{4\pi\,\me c^2}\left(F(x)+G(x)\right);	\\
\Polpar(\om)&=&\frac{\sqrt{3}\,e^3\Bperp}{4\pi\,\me c^2}\left(F(x)-G(x)\right),
\end{eqnarray}
where
\begin{eqnarray}
\label{eq:Fdef}
F(x)&=&x\int_x^\infty{K_{5/3}(z)\,dz};	\\
\label{eq:Gdef}
G(x)&=&xK_{2/3}(x),
\end{eqnarray}
with $K_n(z)$ being a modified Bessel function of the second kind,
\begin{equation}
x=\frac{\om}{\omc}=\frac{2}{3}\frac{\me c}{e\Bperp}\frac{\om}{\gm^2}\hateq\frac{2K}{\Bperp\gm^2},
\end{equation}
(where the constant $K$ is defined by the last equation; hereafter the symbol $\hateq$ will be used to indicate the definition of a new quantity, rather than a result),
and $\Bperp$ the modulus of the projected MF: note that all formulae depend only on the projected MF, while the component radial to the observer does not affect the emission (it could be investigated only through propagation effects, namely through FR).

It is more convenient to express this emission in terms of the Stokes parameters.
To be more precise, by now we consider only their ``local values'', namely their values per unit path; while the actual Stokes parameters (namely those that should match the observations) will be obtained by integration along the line of sight.
The choice of reference orientation for the linear polarization parameters, $Q$ and $U$, is such that:
\begin{eqnarray}
\label{eq:Qorientation}
Q&=&\avg{E_x E_x^*}-\avg{E_y E_y^*};	\\
\label{eq:Uorientation}
U&=&\avg{E_x E_y^*}+\avg{E_y E_x^*},
\end{eqnarray}
where $E_x$ and $E_y$ are the components of the complex amplitude of the electric vector.

Without loss of generality, let us first consider an orientation of axes with the unit vector $\hat x'$ perpendicular to $\Bperp$ and $\hat y'$ parallel to it. In this way, the parameters $\Iloc'$ (total flux) and $\Qloc'$ (difference between the linear polarizations along the two axes) relative to the emission of a single particle are
\begin{eqnarray}
\label{eq:Isingleparticle}
\Iloc'(\om)\!\!\!\!\!&=&\!\!\!\!\!\frac{\Polperp+\Polpar}{4\pi}=\frac{\sqrt{3}\,e^3\Bperp}{8\pi^2\me c^2}F(x)\hateq H\Bperp F(x);	\\
\label{eq:Qsingleparticle}
\Qloc'(\om)\!\!\!\!\!&=&\!\!\!\!\!\frac{\Polperp-\Polpar}{4\pi}=\frac{\sqrt{3}\,e^3\Bperp}{8\pi^2\me c^2}G(x)\hateq H\Bperp G(x),
\end{eqnarray}
where the quantity $H$ is defined by the rightmost equalities. 
Note that the Stokes parameters respectively associated to tilted linear polarization ($\Uloc'$) and circular polarization ($\Vloc'$) are both vanishing.

For any other reference frame $x$-$y$, rotated of an angle $\chi$ (taken to be anti-clockwise) with respect to the $x'$-$y'$ reference frame chosen above, the Stokes parameters transform as
\begin{equation}
\left\{
\begin{array}{lcl}
 \Iloc&=&\Iloc';\\
 \Qloc&=&\;\;\,\cos(2\chi)\Qloc'+\,\sin(2\chi)\Uloc'=\;\;\,\cos(2\chi)\Qloc';\\
 \Uloc&=&-\sin(2\chi)\Qloc'+\,\cos(2\chi)\Uloc'=-\sin(2\chi)\Qloc';\\
 \Vloc&=&\Vloc'=0.
\end{array}
\right.
\label{polariz:transform}
\end{equation}
In terms of the MF components we can write
\begin{equation}
\left\{
\begin{array}{lcl}
 \cos(2\chi)&=&\displaystyle\frac{B_y^2-B_x^2}{\Bperp^2},\\\\
 \sin(2\chi)&=&\displaystyle\frac{2B_xB_y}{\Bperp^2},
\end{array}
\right.
\label{polariz:cossin}
\end{equation}
where $\Bperp^2=B_x^2+B_y^2$.
In the case of a power-law energy distribution
\begin{equation}
n(\gm)=A\gm^{-s}
\label{eq:PLdistribution}
\end{equation}
for the emitting particles, the known formulae for the emissivities (per unit volume) are recovered after integration over $\gm$
\begin{eqnarray}
\label{eq:IlocfPL}
\IlocfPL(\om)&=&\!\!\frac{s+7/3}{s+1}\Wz\Bperp^{(s+1)/2};	\\
\label{eq:QlocfPL}
\QlocfPL(\om)&=&\qquad\;\;\;\Wz\Bperp^{(s+1)/2},
\end{eqnarray}
where
\begin{equation}
\label{eq:Wzdef}
 \Wz=\frac{AH}{4K^{(s-1)/2}}\Gm\left(\frac{s}{4}+\frac{7}{12}\right)\Gm\left(\frac{s}{4}-\frac{1}{12}\right),
\end{equation}
while $\IlocPL$ and $\QlocPL$ are obtained with a rotation of an angle $\chi$, according to Eqs.~\ref{polariz:transform}.
In this case the polarization fraction is
\begin{equation}
\Pi\rs{max}=\frac{\sqrt{\QlocsqPL+\UlocsqPL}}{\IlocPL}=\left|\frac{\QlocfPL}{\IlocfPL}\right|=\frac{s+1}{s+7/3}.
\end{equation}
So far, the distinction between primed and unprimed Stokes parameters has been just formal.
However, starting from the next section, it will become more substantial.
With a MF varying in time the primed reference frame, which is instantaneously oriented with the MF, also changes its orientation with time, and therefore the primed Stokes parameters can only appear, as instantaneous values, inside the integrals that we will use to estimate quantities averaged over fluctuations.

\subsection{Inclusion of an isotropic random MF}
\label{polariz:emiss-randomMF}

\begin{figure*}
\centering
\includegraphics[angle=0,width=8.3truecm]{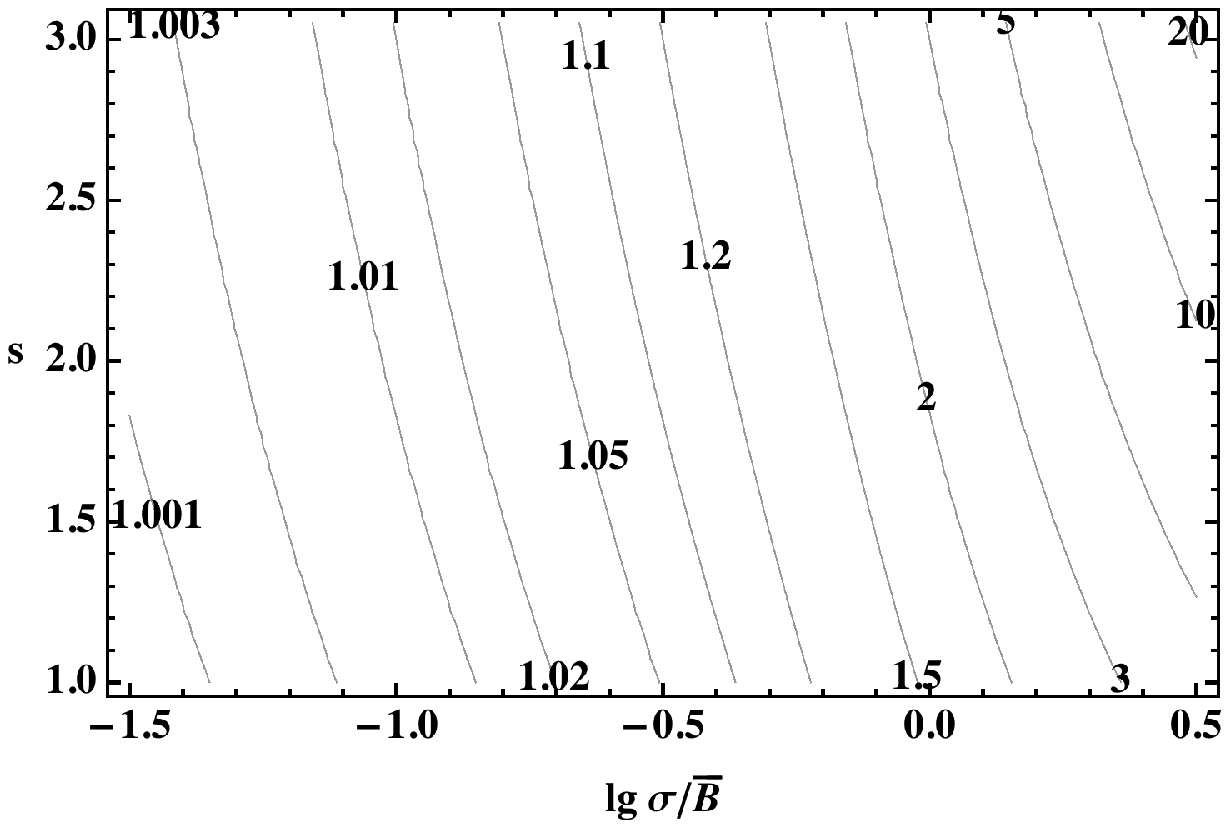}
\includegraphics[angle=0,width=8.3truecm]{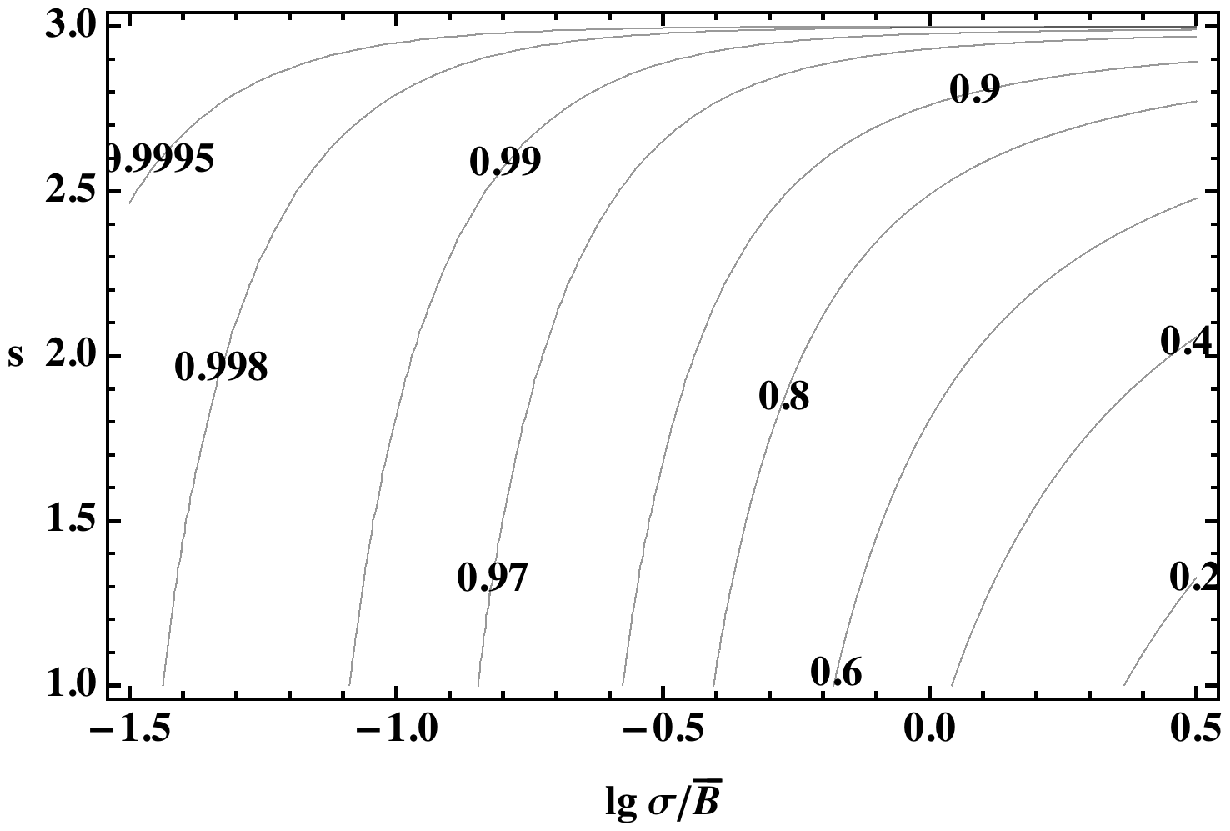}
\caption{
  Maps of $\avgIlocPL/\IlocPL$ (left) and $\avgQlocPL/\QlocPL$ (right), as functions of $\sg/\Bav$ and $s$, in the case in which the random MF component is isotropic.}
\label{fig:IQFractionMap}
\end{figure*}
Scope of this section is to extend the classical treatment of the synchrotron emission to the case in which the MF is a combination of an ordered and of a random component.
While the two limit cases of a uniform MF (maximally anisotropic emissivity, and maximal polarization fraction), and of a completely random MF (isotropic and unpolarized emission) are well known, the intermediate cases are not so obvious.
In particular, it must be clear that in the case of a partly random MF it would be incorrect to estimate the time-averaged synchrotron emissivity as simply the sum of the emissivities of its ordered component and of random part.
We will show, instead, that a general theory can be developed, as a rather natural extension of the treatment for the case of a uniform MF.

Let us consider now a MF $\vectB=\vectBav+\vectdB$, where $\vectBav$ is its average (which, without loss of generality, we assume to be directed along $\hat y$) while $\vectdB$ varies randomly according to an isotropic Gaussian distribution.
Indeed, one may envisage an ample choice of statistical distributions for the random component: for instance, one could have assumed $\vectdB$ to be perpendicular to $\vectBav$.
Our choice is motivated instead not only by an advantage for the computations but also by the fact that, with our recipe, we can treat in a continuous and homogeneous way both cases in which fluctuations are just a small perturbation and those in which they dominate, and that in all cases the observed emission depends uniquely on the properties of the projected MF.
Instead, for instance, the prescription of purely transverse fluctuations would have lead to the paradox that, in the limit of very large fluctuations, an energetically irrelevant $\vectBav$ would still play a leading role, determining the orientation of fluctuations.
Another important constraint on the fluctuations is that they must be consistent with a vanishing divergence of the total MF. However, this constraint applies only to the spatial derivatives of $\vectdB$, while its one-point statistical distribution (as in our treatment) is not affected.

Then, the distribution of the projected components of the combined MF are
\begin{eqnarray}
\Prob_x(B_x)&=&\frac{1}{\sqrt{2\pi}\sg}\exp\left(-\frac{B_x^2}{2\sg^2}\right);	\\
\Prob_y(B_y)&=&\frac{1}{\sqrt{2\pi}\sg}\exp\left(-\frac{(B_y-\Bav)^2}{2\sg^2}\right),
\end{eqnarray}
where $\sg$ is the standard deviation of the fluctuations,  while the projected  component $\Bav$ of the unperturbed MF is taken to be oriented along $y$.
Note that, from now on, we will skip the index $\perp$ on both unperturbed MF and MF perturbations.  
In fact, without loss of generality, we will consider only the transverse MF components.
By changing the element area from $dB_xdB_y$ to $dB\,dB_y$ and then integrating over $B_y$ we obtain
\begin{eqnarray}
\Prob(B)&=&\!\!\!\!\int_{-B}^{+B}{\!\!\!\!\!\!\!\!\!\Prob_x(B_x)\Prob_y(B_y)\,2\left.\frac{\pd B_x}{\pd B}\right|_{B_y}dB_y}	\nonumber	\\
&=&\!\!\!\!\int_{-B}^{+B}{\!\!\!\!\!\!\!\!\!\exp\left(-\frac{B^2-2B_y\Bav+\Bav^2}{2\sg^2}\right)
\frac{B\,dB_y}{\pi\sg^2\sqrt{B^2-B_y^2}}}		\nonumber	\\
&=&\!\!\!\!I_0\!\left(\frac{B\,\Bav}{\sg^2}\right)\exp\left(-\frac{B^2+\Bav^2}{2\sg^2}\right)\frac{B}{\sg^2},
\end{eqnarray}
with $I_n(z)$ being a modified Bessel function of the first kind (note that during that change of variable a factor 2 has been introduced to keep track of the multiplicity in $B_x$).

The average Stokes parameters in the case of a fluctuating MF can be obtained by averaging over the probability distribution of the MF fluctuations.
For the total intensity $\Iloc$ we then simply have
\begin{eqnarray}
\avg{\Iloc}\!&=&\int_0^\infty{\Iloc\,\Prob(B)\,dB}	\nonumber	\\
\label{eq:Iavsingleparticle}
&\!\!\!\!\!\!\!\!\!\!\!\!\!\!\!\!\!\!\!\!\!\!\!\!\!\!=&\!\!\!\!\!\!\!\!\!\!\!\!\!\!\!\!\!\!\!\!
\int_0^\infty{\!\!\!HF\!\left(\frac{2K}{B\gm^2}\right)I_0\!\left(\frac{B\,\Bav}{\sg^2}\right)\exp\left(-\frac{B^2+\Bav^2}{2\sg^2}\right)\frac{B^2dB}{\sg^2}}.
\end{eqnarray}
As for the Stokes parameters $Q$ and $U$, one must note that, while the average MF is oriented along $y$, this is not the case for the instantaneous MF, in which case these parameters (given above with respect of the orientation of the MF) must all be derotated to the $y$ axis, by using Eqs.~8 and 9, before integration.
We then obtain
\begin{eqnarray}
\avg{\Qloc}\!\!\!\!\!&=&\!\!\!\!\!\!\int{\!\!\!HG\!\left(\frac{2K}{B\gm^2}\right)\exp\left(-\frac{B^2-2B_y\Bav+\Bav^2}{2\sg^2}\right)}\nonumber	\\
\label{eq:Qavsingleparticle}
&&\qquad\qquad\qquad\qquad\cdot\frac{(B^2-2B_y^2)\,dB_y}{\pi\sg^2\sqrt{B_y^2}}dB\\
&=&\!\!\!\!\!\!\int{\!\!\!HG\!\left(\frac{2K}{B\gm^2}\right)I_2\!\left(\frac{B\,\Bav}{\sg^2}\right)\exp\left(-\frac{B^2+\Bav^2}{2\sg^2}\right)\!\frac{B^2dB}{\sg^2}};	\nonumber
\end{eqnarray}
while $\avg{\Uloc}=0$ for symmetry reasons.
All the above formulae refer to an ensemble of particles with a fixed Lorentz factor $\gm$, while in the case of a general energy distribution function one should convolve the above expressions for $\avg{\Iloc}$ and  $\avg{\Qloc}$ by the particle distribution.

\begin{figure}
\centering
\includegraphics[angle=0,width=8.3truecm]{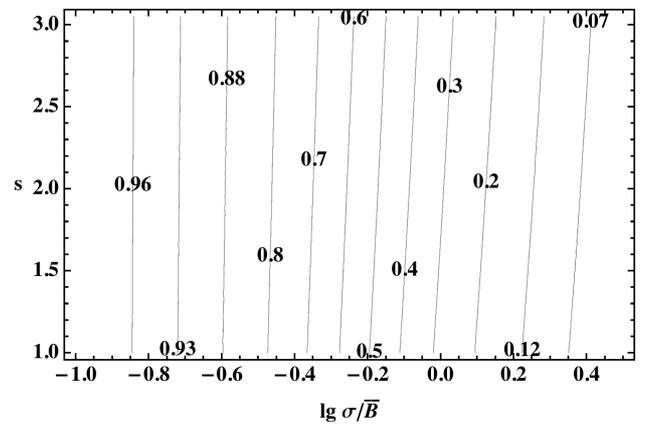}
\caption{Map of $\Pi/\Pimax$ for isotropic fluctuations, where $\Pimax$ is the case for a uniform MF. It is worth noting how weakly does $\Pi/\Pimax$ depend on $s$.}
\label{fig:PolFractionMapB}
\end{figure}
\begin{figure*}
\centering
\includegraphics[angle=0,width=8.3truecm]{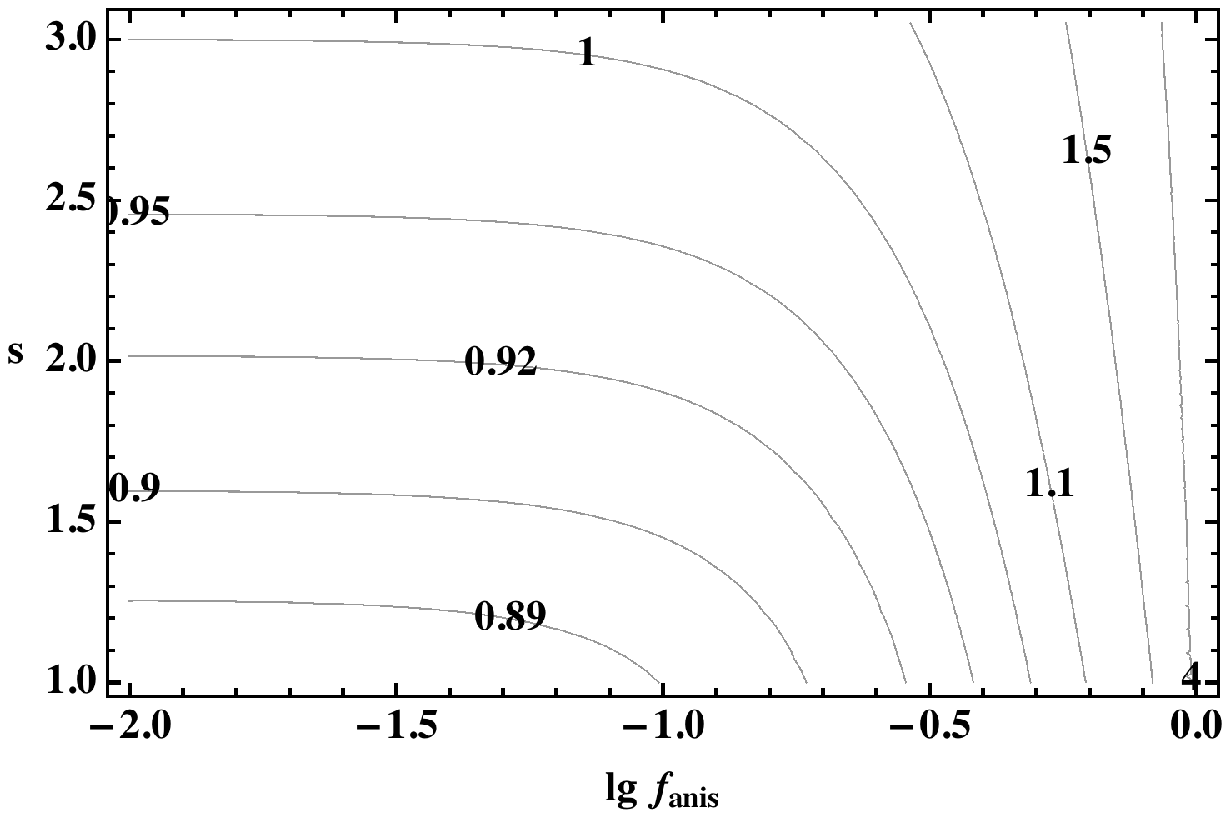}
\includegraphics[angle=0,width=8.3truecm]{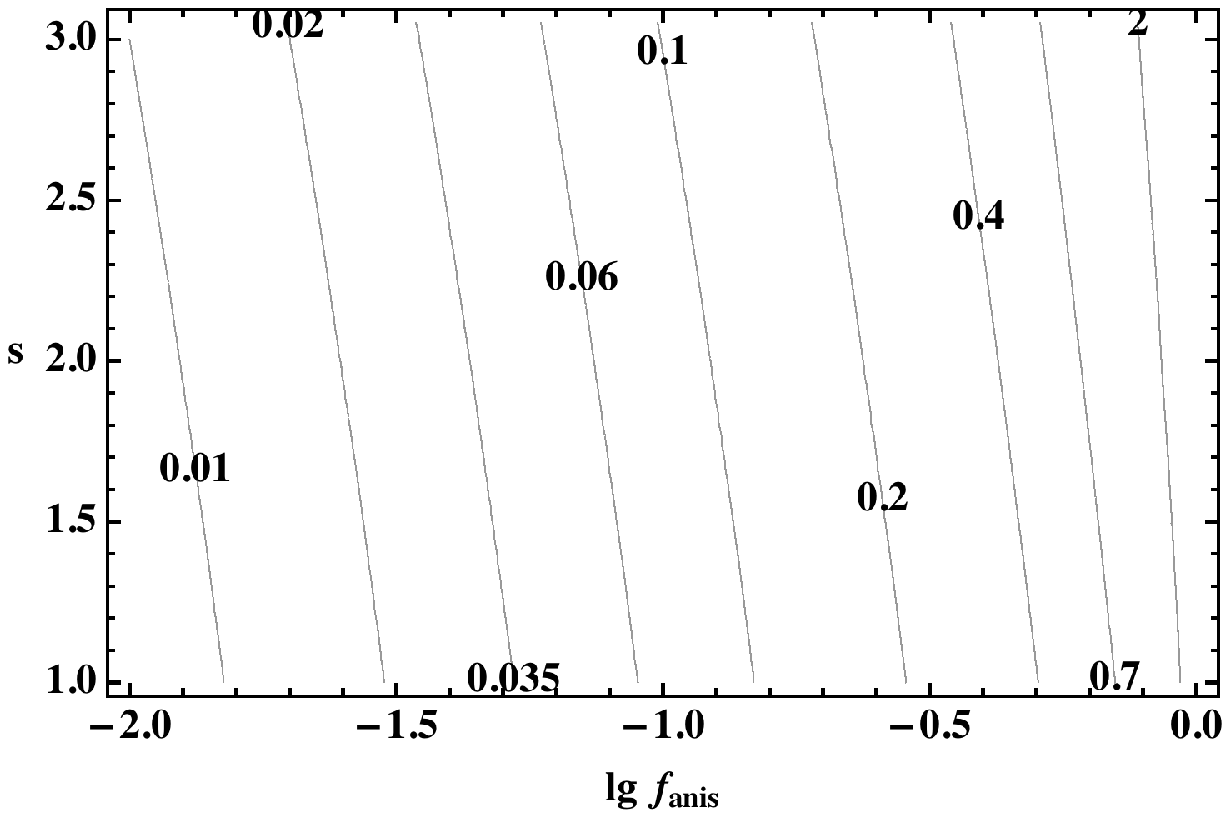}
\caption{
Maps of $\avgIlocPL/\IlocPL$ and $\avgQlocPL/\QlocPL$, in the case of anisotropic fluctuations and a vanishing ordered MF component, as functions of $\fanis$ and $s$.}
\label{fig:IQFractionMapanis}
\end{figure*}

This convolution can be performed rather easily in the case of a pure power-law energy distribution $n(\gm)=A\gm^{-s}$.
In this case, by making use of the following identities,
\begin{eqnarray}
\int_0^\infty{x^\mu F(x)\,dx}&=&\frac{2^{\mu+1}}{\mu+2}\opGm\left(\frac{\mu}{2}+\frac{7}{3}\right)\opGm\left(\frac{\mu}{2}+\frac{2}{3}\right);	\\
\int_0^\infty{x^\mu G(x)\,dx}&=&2^\mu\opGm\left(\frac{\mu}{2}+\frac{4}{3}\right)\opGm\left(\frac{\mu}{2}+\frac{2}{3}\right),
\end{eqnarray}
one gets
\begin{eqnarray}
\avgIlocPL\!\!\!\!\!\!&=&\!\!\!\!\!\!\frac{s+7/3}{s+1}\Wz\sg^{(s+1)/2}\int{I_0\left(\frac{B\,\Bav}{\sg^2}\right)}	\nonumber	\\
&&\qquad\qquad	\cdot\exp\left(-\frac{B^2+\Bav^2}{2\sg^2}\right)\frac{B^{(s+3)/2}dB}{\sg^{(s+5)/2}};	\\
\avgQlocPL\!\!\!\!\!\!&=&\qquad\;\Wz\sg^{(s+1)/2}\int{I_2\left(\frac{B\,\Bav}{\sg^2}\right)}	\nonumber	\\
&&\qquad\qquad	\cdot\exp\left(-\frac{B^2+\Bav^2}{2\sg^2}\right)\frac{B^{(s+3)/2}dB}{\sg^{(s+5)/2}}.
\end{eqnarray}
The integrations can be performed in terms of special functions, giving
\begin{eqnarray}
\avgIlocPL\!\!\!\!\!\!&=&\!\!\!\!\!\!\IlocPL\left\{\Gm\left(\frac{5+s}{4}\right)\left(\frac{\Bav}{\sqrt{2}\sg}\right)^{-(1+s)/2}
\right.\nonumber	\\
\label{eq:IPL}
&&\qquad\qquad	\cdot\left.
{_{1}F_{1}}\!\left(-\frac{1+s}{4},1,-\frac{\Bav^2}{2\sg^2}\right)\right\}
;	\\
\avgQlocPL\!\!\!\!\!\!&=&\!\!\!\!\!\!\QlocPL\left\{\frac{1}{2}\Gm\left(\frac{9+s}{4}\right)\left(\frac{\Bav}{\sqrt{2}\sg}\right)^{(3-s)/2}
\right.\nonumber	\\
\label{eq:QPL}
&&\qquad\qquad	\cdot\left.
{_{1}F_{1}}\!\left(\frac{3-s}{4},3,-\frac{\Bav^2}{2\sg^2}\right)\right\},
\end{eqnarray}
where ${_{1}F_{1}}(a,b,z)$ is the Kummer confluent hypergeometric function.
Fig.~\ref{fig:IQFractionMap} shows that both factors in braces approach unity for $\sg/\Bav\rightarrow 0$ and the known results are recovered for a vanishing random component of MF.
In the opposite limit, both hypergeometric factors approach unity, so that asymptotically
\begin{eqnarray}
\avgIlocPLasymp&=&\!\!\!\frac{s+7/3}{s+1}\Gm\left(\frac{5+s}{4}\right)\Wz\left(\sqrt{2}\sg\right)^{(s+1)/2}, \\
\avgQlocPLasymp&=&\frac{1}{2}\Gm\left(\frac{9+s}{4}\right)\Wz\Bav^2\left(\sqrt{2}\sg\right)^{(s-3)/2}
\end{eqnarray}
(cf. Eqs.~\ref{eq:IlocfPL} and \ref{eq:QlocfPL}).
This means that, for $\sg\gg\Bav$, apart from a numerical factor the asymptotic formula for $\avgIlocPL$ is formally similar to $\IlocPL$, with $\sqrt{2}\sg$ taking the place of the ordered MF; while the formula for $\avgQlocPL$ is asymptotically proportional to $\Bav^2\sg^{(s-3)/2}$, leading to the asymptotic formula for the polarization fraction
\begin{equation}
\Pi\rs{asymp}\simeq\frac{s+1}{s+7/3}\frac{5+s}{8}\frac{\Bav^2}{2\sg^2}
\end{equation}
from which one can easily see how the polarization fraction vanishes, in the limit of a large random MF component.

To conclude, the general formula for the polarization fraction becomes
\begin{equation}
\Pi\!=\!\Pimax\frac{(5+s)}{8}\frac{\Bav^2}{2\sg^2}\frac{{_{1}F_{1}}\!\left((3-s)/4,3,-\Bav^2/2\sg^2\right)}{{_{1}F_{1}}\!\left(-(
1+s)/4,1,-\Bav^2/2\sg^2\right)}.
\end{equation}
The general behaviour of $\Pi/\Pimax$ is given in Fig.~\ref{fig:PolFractionMapB}, showing that $\Pi/\Pimax$ is only weakly dependent on $s$.

\subsection{An anisotropic random MF}
\label{polariz:emiss-anisrandomMF}

A scenario somehow different from that outlined above applies to the case of efficient particle acceleration in a quasi-parallel shock \citep[see e.g.][and references therein]{Caprioli-Spitkovsky-2014a}: in this case a strong MF turbulent amplification takes place upstream of the shock front.
This turbulent MF is roughly isotropic, as long as it keeps upstream; but the compression at the shock has the effect of enhancing the field components parallel to the shock front, leading to an anisotropic random MF component downstream.
Under general conditions, this case is more complex than that outlined in the previous section, and does not allow an analytic solution.
However, it becomes considerably simpler in the case of a negligible original MF ($\Bav=0$): a condition physically rather reasonable, whenever the MF experiences a strong amplification.

Let us consider also in this case just the projected MF, and assume the following distributions for the two components
\begin{eqnarray}
\Prob_x(B_x)&=&\frac{1}{\sqrt{2\pi}\sg_x}\exp\left(-\frac{B_x^2}{2\sg_x^2}\right);	\\
\Prob_y(B_y)&=&\frac{1}{\sqrt{2\pi}\sg_y}\exp\left(-\frac{B_y^2}{2\sg_y^2}\right),
\end{eqnarray}
where $\sg_x$ and $\sg_y$ are the standard deviations of the fluctuations along the two axes (here we also assume $\sg_y>\sg_x$; while the more general case can be obtained by a mere axes rotation).
In an analogous way to the previous section
\begin{eqnarray}
\Prob(B)&=&\!\!\!\!\int_{-B}^{+B}{\!\!\!\!\!\!\!\!\!\exp\left(-\frac{B^2-B_y^2}{2\sg_x^2}-\frac{B_y^2}{2\sg_y^2}\right)
\frac{B\,dB_y}{\pi\sg_x\sg_y\sqrt{B^2-B_y^2}}}		\nonumber	\\
&=&\!\!\!\!I_0\!\left(B^2\frac{\sg_y^2-\sg_x^2}{4\sg_x^2\sg_y^2}\!\right)\exp\left(\!-B^2\frac{\sg_x^2+\sg_y^2}{4\sg_x^2\sg_y^2}\!\right)\frac{B}{\sg_x\sg_y}.
\end{eqnarray}
The average Stokes parameters can be then evaluated as
\begin{eqnarray}
\label{eq:Ianis1}
\avg{\Iloc}&=&
\int_0^\infty{HF\left(\frac{2K}{B\gm^2}\right)I_0\left(B^2\frac{\sg_y^2-\sg_x^2}{4\sg_x^2\sg_y^2}\right)}
\nonumber \\
&&\qquad\qquad\cdot\exp\left(-B^2\frac{\sg_x^2+\sg_y^2}{4\sg_x^2\sg_y^2}\right)\frac{B^2dB}{\sg_x\sg_y}; \\
\label{eq:Qanis1}
\avg{\Qloc}&=&
\int_0^\infty{HG\left(\frac{2K}{B\gm^2}\right)I_1\left(B^2\frac{\sg_y^2-\sg_x^2}{4\sg_x^2\sg_y^2}\right)}
\nonumber \\
&&\qquad\qquad\cdot\exp\left(-B^2\frac{\sg_x^2+\sg_y^2}{4\sg_x^2\sg_y^2}\right)\frac{B^2dB}{\sg_x\sg_y},
\end{eqnarray}
while, again, $\avg{\Uloc}=0$ for symmetry reasons.
These relations can be simplified, by introducing an effective dispersion and an asymmetry factor, defined respectively as
\begin{equation}
\sgeff^2=\frac{2\sg_x^2\sg_y^2}{\sg_x^2+\sg_y^2};\qquad
\fanis=\frac{\sg_y^2-\sg_x^2}{\sg_x^2+\sg_y^2}.
\end{equation}
Eqs.~\ref{eq:Ianis1} and \ref{eq:Qanis1} can be then rewritten as

\begin{figure}
 \centering
\includegraphics[angle=0,width=8.3truecm]{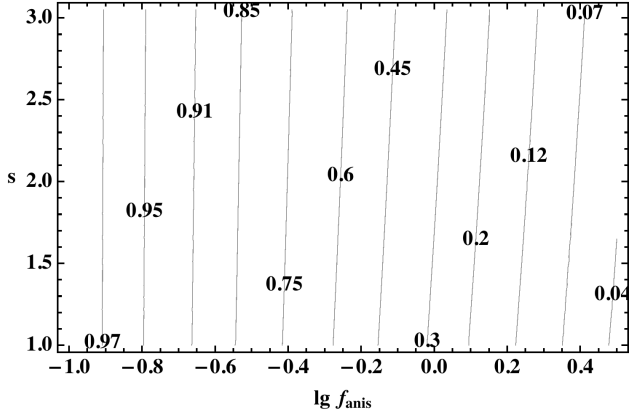}
\caption{Map of $\Pi/\Pimax$ in the case of anisotropic fluctuations and a vanishing ordered MF component. Note that also} in this case $\Pi/\Pimax$ depends very weakly on $s$.
\label{fig:PolFractionMapBanis}
\end{figure}

\begin{figure}
 \centering
\includegraphics[angle=0,width=8.3truecm]{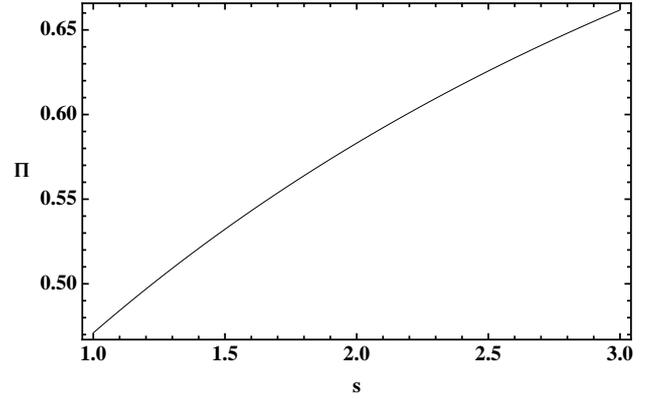}
\caption{Polarization fraction as a function of $s$, in the case of a shock compression of an originally isotropic random MF.}
\label{fig:PolarizedShock}
\end{figure}

\begin{eqnarray}
\avg{\Iloc}&=&
\int_0^\infty{HF\left(\frac{2K}{B\gm^2}\right)I_0\left(\fanis\frac{B^2}{2\sgeff^2}\right)} \nonumber	\\
\label{eq:Ianis2}
&&\qquad\qquad\cdot\exp\left(-\frac{B^2}{2\sgeff^2}\right)\frac{\sqrt{1-\fanis^2}B^2dB}{\sgeff^2}; 	\\
\avg{\Qloc}&=&
\int_0^\infty{HG\left(\frac{2K}{B\gm^2}\right)I_1\left(\fanis\frac{B^2}{2\sgeff^2}\right) }\nonumber	\\
\label{eq:Qanis2}
&&\qquad\qquad\cdot\exp\left(-\frac{B^2}{2\sgeff^2}\right)\frac{\sqrt{1-\fanis^2}B^2dB}{\sgeff^2}.
\end{eqnarray}
The convolution with a pure power-law energy distribution finally leads to
\begin{eqnarray}
\avgIlocPL&=&\left(\frac{s+7/3}{s+1}\Wz(2\sgeff^2)^{(s+1)/4}\right)	\nonumber \\
\label{eq:IPL2}
&&\!\!\!\!\!\!\!\!\!\!\!\!\!\!\!\!\!\!\!\!\!\!\!\!\cdot\left\{\sqrt{1-\fanis^2}\,\Gm\left(\frac{5+s}{4}\right){_{2}F_{1}}\!\left(\frac{5+s}{8},\frac{9+s}{8},1,\fanis^2\right)\right\};	\\
\avgQlocPL&=&\left(\Wz(2\sgeff^2)^{(s+1)/4}\right)	\nonumber \\
\label{eq:QPL2}
&&\!\!\!\!\!\!\!\!\!\!\!\!\!\!\!\!\!\!\!\!\!\!\!\!\!\!\!\!\!\!\!\!\!\!\cdot\left\{\!\frac{\fanis}{2}\sqrt{1-\fanis^2}\,\Gm\left(\frac{9+s}{4}\right){_{2}F_{1}}\!\left(\frac{9+s}{8},\frac{13+s}{8},2,\fanis^2\right)\!\right\},
\end{eqnarray}
where ${_{2}F_{1}}(a,b,c,z)$ is the Gauss hypergeometric function
(see Fig.~\ref{fig:IQFractionMapanis}). It is worth noting, in the two above equations, the formal similarity of the leading factors and Eqs.~\ref{eq:IlocfPL}, \ref{eq:QlocfPL}, with $\sqrt{2}\sgeff$ taking the place of $\Bperp$.
The general formula for the polarization fraction (see Fig.~\ref{fig:IQFractionMapanis}) then becomes
\begin{equation}
\Pi=\Pimax
\cdot\left\{\frac{(5+s)\fanis\,\,{_{2}F_{1}}\!\left((9+s)/8,(13+s)/8,2,\fanis^2\right)}{8\,\,{_{2}F_{1}}\!\left((5+s)/8,(9+s)/8,1,\fanis^2\right)}\right\}.
\label{eq:PolarAnis}
\end{equation}
The ratio $\Pi/\Pimax$, as shown in Fig.~\ref{fig:PolFractionMapBanis}, is only weakly dependent on $s$.
In the limit of $\sg_x\simeq\sg_y$, (i.e. $\sg=(\sg_x+\sg_y)/2$ and $\dl\sg=\sg_y-\sg_x\ll\sg$)
\begin{equation}
\sgeff\simeq\sg-\frac{3\,\dl\sg^2}{8\sg};\qquad \fanis\simeq\frac{\dl\sg}{\sg},
\end{equation}
and then the factors in braces in Eqs.~\ref{eq:IPL2} and \ref{eq:QPL2} respectively approach
\begin{equation}
\Gm\left(\frac{5+s}{4}\right)\qquad\hbox{and}\quad
\frac{5+s}{4}\,\Gm\left(\frac{5+s}{4}\right)\frac{\fanis}{2},
\end{equation}
then leading to a small polarization fraction
\begin{equation}
\Pi=\frac{s+1}{s+7/3}\frac{5+s}{8}\fanis.
\end{equation}
Instead, in the limit $\sg_y \ll\sg_x$
\begin{equation}
\sgeff\simeq\sqrt{2}\sg_y;\qquad \fanis\simeq1-\frac{2\sg_y^2}{\sg_x^2},
\end{equation}
and in this case both factors within braces in Eqs.~\ref{eq:IPL2} and \ref{eq:QPL2} approach
\begin{equation}
\Gm\left(\frac{3+s}{4}\right)\frac{\left(1-\fanis\right)^{-(1+s)/4}}{\sqrt{\pi}},
\end{equation}
therefore leading, as expected, to the same limit value of the polarization fraction as in the case of a fully ordered MF.

The above results can be readily applied to estimate the polarization induced by a shock compression.
If the projected shock velocity is along the $x$ axis, in the downstream $\sg_x$ would be equal to $\sg$ in the upstream, while $\sg_y$ would be enhanced by a factor $\comprratio$ (where $\comprratio$ is the shock compression ratio).
In the case of a strong shock with $\comprratio=4$, then, $\fanis$ would be equal to $15/17$; the related polarization fraction, evaluated from Eq.~\ref{eq:PolarAnis} is shown in Fig.~\ref{fig:PolarizedShock}. It is worth noting that in this case the downstream emission still exhibits a strong polarization, much stronger than typically measured in typical shell-like SNRs.

\subsection{Mono-energetic electron distribution}
\label{polariz:deltadistribution}

We have shown how, in the case of a power-law energy distribution of the electrons, one may obtain in a rather elegant way analytic formulae for the polarized emission that resemble the classical formulae for a the case of a fully ordered MF.

Unfortunately, for a general particle energy distribution the situation is much more complex, and there is no guarantee about the existence of analytic solutions.
Just to give an feeling of the kind of additional problems arising in the general case, let us consider here the case of total intensity ($\avg{\Iloc}_{\gm}$) for a mono-energetic distribution of particles: this case requires one to solve Eq.~\ref{eq:Iavsingleparticle}, for a fixed $\gm$ value.
In spite of the expected simplicity of this problem, we have not found any analytic solution to it except for the two limiting cases, namely a fully ordered and a completely random MF.
\begin{figure}
\centering
\includegraphics[angle=0,width=8.3truecm]{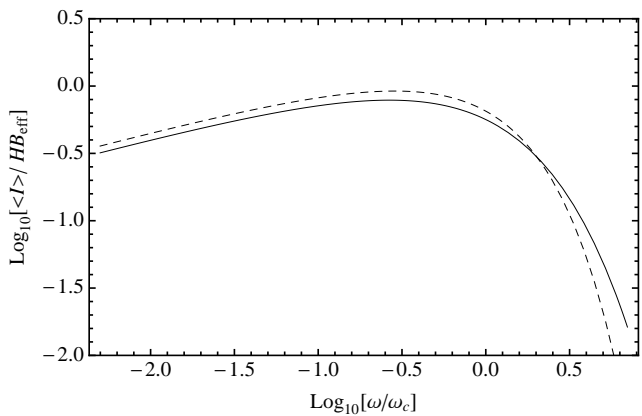}
\caption{Comparison of the spectral profile of the synchrotron emissivity from a mono-energetic particles distribution, both in the case of a completely ordered MF (dashed line), and in that of a fully random MF (solid line).
$B\rs{eff}$ indicates the effective MF for each case, namely $\Bav$ in the former case, and $\sqrt{2}\sg$ in the latter one.}
\label{fig:CompareSingleEnergy}
\end{figure}

The former case is simply proportional to the emitted spectrum from a single particle, Eq.~\ref{eq:Isingleparticle}.
The latter case, instead, is obtained by evaluating Eq.~\ref{eq:Iavsingleparticle} in the limit $\Bav=0$.
Using Eq.~\ref{eq:Fdef} for the definition of $F(x)$, and inverting the integration order, one may reduce Eq.~\ref{eq:Iavsingleparticle} to
\begin{eqnarray}
\avg{\Iloc}&=&H\sg\int_0^\infty{K_{5/3}(z)\frac{\om}{\omcsg}\exp\left(-\frac{\om^2}{\omcsg^2}\frac{1}{2z^2}\right)\,dz}	\nonumber	\\
&=&H\sg\frac{\pi}{\sqrt{3}}\left(\frac{\om}{\omcsg}+\frac{4}{3}\frac{\om^{1/3}}{\omcsg^{1/3}}\right)\exp\left(-\frac{3}{2}\frac{\om^{2/3}}{\omcsg^{2/3}}\right)	\\
&\hateq&H\sg F_\sg(\om/\omcsg), \nonumber
\end{eqnarray}
where we have introduced the quantities $F_\sg$ (defined above), and
\begin{equation}
\omcsg=\frac{3}{2}\frac{e\,\sg}{\me c}\gm^2,
\end{equation}
in order to obtain a formula that formally resembles Eq.~\ref{eq:Isingleparticle}.
Fig.~\ref{fig:CompareSingleEnergy} shows a comparison between $F(x)$ and $F_\sg(\sqrt{2}x)/\sqrt{2}$ (where the $\sqrt{2}$ factors come from the fact that, in the completely random case, the effective MF strength would be equal to $\sqrt{2}\sg$).
The present analysis is in a sense similar to that in \citet{Pohl-etal-2015}; the main differences are that that paper assumes a 1-D Gaussian distribution for the MF values and an approximation of $F(x)$, while here we use a 2-D one, which we think is more correct, and the exact $F(x)$ function: the interesting fact is that our approach luckily allows an exact solution.

However, no analytic solution is known for cases intermediate between a fully ordered and a fully random field. This issue, which is also preliminary to any further generalization to particle energy distributions different from the pure power-law one (the case thoroughly treated in the present work), will need a separate treatment.

\section{Modelling shell-type SNRs}
\label{polariz:models}

In order to apply the formulae derived in the previous section and to synthesize the emission and polarization maps for shell-like SNRs, we need to introduce SNR models on which to integrate the contribution of all volume elements along the lines of sight.
In the most general case the number of parameters would be unfortunately so large to hinder any extensive investigation in the parameters space.

For this reason in the following of this paper we will limit our discussion to the case of a thin layer model of a spherical shell-type SNR.
According to this approximation, in each position of the map the observed emission originates from two (geometrically thin) limbs, at a physical distance $\Rs$ from the SNR center.
Moreover, we assume that we can substitute the actual profiles of all physical quantities (like the density of the thermal gas, that of the relativistic particles, as well as the MF direction and strength) with their average values.
As we shall see, also this very simplified approximation my lead to a rather complex and various phenomenology.
In a forthcoming paper we shall release this assumption, and discuss the validity and limits of the thin layer approximation.

In the rest of this section we will describe in more detail the various elements of this model.

\subsection{Physical and projected coordinates}

A very important although just geometrical task is how to convert specific elements in the SNR from/to points on the observed map, and in this section we present some relations that allow us to make this conversion easier.

Let $\{\rad,\tht,\varphi\}$ be the spherical coordinates in a reference system oriented with respect to a given axis that we will identify with the SNR axis.
In order to further simplify the modelling, we also assume that the SNR is axially symmetric, namely no explicit dependence on $\varphi$ of the SNR physical quantities.
Moreover, as before, let $\{x,y\}$ be the coordinates of the map and $z$ that along the line of sight.

Let us also define $\radproj$ and $\Tht$ as the projected radial and angular coordinates, and $\phiz$ as the angle (aspect angle) by which the SNR axis is tilted with respect to the line of sight.
The projection of the SNR axis can be assumed, without loss of generality, to lie along the $x$ direction.
A preliminary step is to set the conversion between the two sets of coordinates.
Under the above assumptions, the map coordinates read
\begin{eqnarray}
\label{eq:xformula}
x&\!\!\!\!=\!\!\!\!&\radproj\cos\Tht=\rad\left(\cosphiz\sintht\cosphi+\sinphiz\costht\right);	\\ 
\label{eq:yformula}
y&\!\!\!\!=\!\!\!\!&\radproj\sin\Tht=\rad\;\sintht\sinphi,
\end{eqnarray}
while the line of sight coordinate
\begin{equation}
\label{eq:zformula}
z\quad=\;\;\rad\left(\cosphiz\costht-\sinphiz\sintht\cosphi\right).
\end{equation}
Using Eqs.~\ref{eq:xformula} and \ref{eq:yformula} one can derive
\begin{eqnarray}
\label{eq:cosphi}
\cosphi&=&\frac{\tradproj\cosTht-\sinphiz\costht}{\cosphiz\sintht};	\\ 
\label{eq:sinphi}
\sinphi&=&\frac{\tradproj\sinTht}{\sintht},
\end{eqnarray}
where $\tradproj=\radproj/\rad$, which is always less than unity.
It is worth noting that, if we consider a spherical shell, the tilt angle ($\tau$) of an element of surface would be such that $\sintau=\tradproj$.

Eqs.~\ref{eq:cosphi} and \ref{eq:sinphi} can be combined to give
\begin{equation}
\costht=\tradproj\sinphiz\cosTht\pm\cosphiz\sqrt{1-\tradproj^2},
\end{equation}
the two solutions referring respectively to the nearer and further limb of the sphere of radius $\rad$, intercepted by the line of sight.

\subsection{MF geometry}

Probably the most natural geometry for the ordered MF component is that based in the idea that the SNR expands in an ambient medium with a pre-existing uniform ambient MF $\Bamb$, and that the MF structure inside the SNR itself is a direct consequence of compression.
We also assume that the morphology to the SNR emission is affected by this MF geometry, and in particular that the SNR symmetry axis is aligned with the ambient MF direction.

Within the thin-layer approach, all MF lines are asymptotically aligned along the meridians.
Therefore, in the SNR reference frame, the magnetic unit vector is
\begin{equation}
\unitBav=\left\{-\costht\cosphi,-\costht\sinphi,\sintht\right\},
\label{eq:uvBmeridional}
\end{equation}
or with opposite sign, depending on the MF orientation (anyway, this orientation does not affect in any way the values of the Stokes parameters).
The projection of this unit vector on the $x$-$y$ plane is then
\begin{equation}
\left\{\frac{\sinphiz-\tradproj\cosTht\costht}{\sintht},-\frac{\tradproj\sinTht\costht}{\sintht}\right\};
\end{equation}
while the magnetic pitch angle $\al$ is
\begin{equation}
\cosal=\frac{\sintht^2-\sinphiz^2+\tradproj\cosTht\costht\sinphiz}{\cosphiz\sintht}.
\end{equation}
As for the strength of the ordered MF, the flux conservation leads to
\begin{equation}
\Bav=\frac{\sintht\,\Bamb}{1-(1-w(\tht)/\Rs)^2}\simeq\frac{\sintht\,\Bamb}{2w(\tht)/\Rs},
\label{eq:magBmeridional}
\end{equation}
where $w(\tht)$ is the layer thickness, the thin-layer approximation requiring $w(\tht)\ll\Rs$ at all $\tht$.
This results in an artificially large B, but it does not affect the patterns of polarization maps.

Also the densities of thermal and relativistic particle components will depend on the layer thickness.
The mass conservation implies:
\begin{equation}
\rho=\frac{\rhoamb}{1-(1-w(\tht)/\Rs)^3}\simeq\frac{\rhoamb}{3\,w(\tht)/\Rs}.
\end{equation}
Let us also assume that the relativistic component is frozen in the MF and evolves adiabatically (the Lorentz factors then evolving as $\gm\propto\rho^{1/3}$).
In this case the normalization factor $A$, as defined in Eq.~\ref{eq:PLdistribution}, is
\begin{equation}
\label{eq:Aevol}
A=\As\left(\frac{\rho}{\rhos}\right)^{1+(s-1)/3}=\As(\tht)\left(\frac{3w(\tht)}{4\Rs}\right)^{(s+2)/3}.
\end{equation}
The quantity $\As(\tht)$, namely its value right downstream of the shock, is generally unknown.
Let us assume here that, for the sake of simplicity, both $w(\tht)$ and $\As(\tht)$ are constant.
These values are not important here, because in the rest of the paper we will concentrate on the form factors, and we will present only normalized emission maps.
Let us add that, as shown by Eqs.~\ref{eq:IlocfPL}--\ref{eq:Wzdef}, the synchrotron emissivity at a given frequency $\om$ is proportional to $A\Bperp^{(s+1)/2}$.
Therefore there is a degeneracy between the layer thickness, the normalization of the energy distribution of particles, and the MF strength (to the power $(s+1)/2$), in the sense that if we scale any of them by a given factor the resulting emissivity will be scaled by the same factor, and leaving completely unchanged the pattern of the various maps.

Another MF geometry, suggested by radio polarization maps of younger SNRs \citep[][]{Milne-1987}, is the radial one, namely
\begin{equation}
\unitBav=\left\{\sintht\cosphi,\sintht\sinphi,\costht\right\}.
\end{equation}
Of course, this does not imply the (quite unphysical) assumption of a magnetic monopole.
Since the polarization properties of the synchrotron emission do not depend on the MF orientation, one could imagine a mixture of inward and outward oriented radial MF lines.
Differently from the previous case, now there is no obvious recipe about the dependence of the ordered MF strength on the position, so we will have to choose it {\sl a priori}
(see the two cases discussed in Sect.~5).

\subsection{Faraday Rotation}

The Stokes parameters projected onto the plane of the sky can be obtained by integration of the local values along the line of sight.
However, for the two Stokes parameters related to linear polarization, one should also take into account the FR, a propagation effect that produces a rotation (by an angle $\bt$) of the polarization plane, according to the formula
\begin{equation}
\bt(z_1,z_2)=\frac{e^3\lmb^2}{2\pi\me^2c^4}\int_{z_1}^{z_2}n(z')\Bpar(z')\,dz',
\label{eq:polbetadef}
\end{equation}
with $n$ being the local plasma electron density, $\lmb=2\pi c/\om$ the radiation wavelength, and $z_2$, $z_1$ two given positions along the line of sight (of which $z_2$ is the closer to the observer).
By defining the rotation measure as
\begin{equation}
\RM(z_1,z_2)=\frac{e^3}{2\pi\me^2c^4}\int_{z_1}^{z_2}n(z')\Bpar(z')\,dz',
\label{eq:polarizRMdef}
\end{equation}
one can simply write $\bt=\RM\,\lmb^2$.
In order to simplify the notation in some points, we will also use $\RM(z)=\RM(z,\zobs)$ (where $\zobs$ is at the observer's location).
This formula for the FR is linear in $B$, which means that only the ordered MF matters, while the effect of fluctuations averages to zero.

It is however necessary to distinguish between ``foreground FR'' (namely that produced by the medium between the SNR and the observer) and ``internal FR'' (namely that produced inside the SNR itself).
The effect of a foreground medium on the observed Stokes parameters (of course, only $Q$ and $U$ are affected) is the following
\begin{eqnarray}
\label{eq:standardQFR}
Q&=&\cos(2\bt)\Qi-\sin(2\bt)\Ui;	\\ 
\label{eq:standardUFR}
U&=&\sin(2\bt)\Qi+\cos(2\bt)\Ui,
\end{eqnarray}
where $\Qi$ and $\Ui$ indicate the intrinsic Stokes parameters, namely in the absence of FR, as it is the case for $\lmb\rightarrow0$; while $\bt$ is the angle of rotation accumulated between the source boundary closest to the observer, and $\zobs$.
It is worth noting that the above formulae are not only valid to describe the effect of the interstellar medium in front of the SNR, but also to that of the SNR front-side limb with respect to the emission from the rear-side limb.

Instead, the effect of the ``internal FR'' on the observed polarization is more complex, since different rotation measures apply to the radiation emitted in different layers, so that a double integral is required to evaluate its overall effect.

As it will better shown below, a more general relation between the intrinsic and observed Stokes parameters can be described by the relations
\begin{eqnarray}
\label{eq:observedQFR}
Q\!\!\!\!\!&=&\!\!\!\!\!\fobs\Big(\cos(2\RMobs\lmb^2)\Qi-\sin(2\RMobs\lmb^2)\Ui\Big);	\\ 
\label{eq:observedUFR}
U\!\!\!\!\!&=&\!\!\!\!\!\fobs\Big(\sin(2\RMobs\lmb^2)\Qi+\cos(2\RMobs\lmb^2)\Ui\Big),
\end{eqnarray}
which allow us define what we intend by ``observed'' Faraday depolarization factor ($\fobs$) and RM ($\RMobs$).
Only in the case of a purely foreground FR we have $\fobs=1$ and $\RMobs$ equal to the quantity described by Eq.~\ref{eq:polarizRMdef}; while in the general case both $\fobs$ and $\RMobs$ will depend on $\lmb$.

Finally, it is worth noting that, while the FR due to a foreground medium could be completely corrected in the post-processing phase (provided that the radio data have a sufficiently narrow bandwidth), no such correction could be applied in the case of internal FR, since the intrinsic depolarization is irreversible.

\subsection{Observed polarization}

Before proceeding with the derivation of the observed Stokes parameters, let us draw there the general equations that show how they are obtained, by integrating along the line of sight, and also accounting for the FR effect:
\begin{equation}
 \left\{
\begin{array}{lcl}
  I  &=&\int_{z_1}^{z_2}{\cal I}(z')dz';\\ \\
  Q&=&\int_{z_1}^{z_2}\Big(\cos\left(2\bt(z_1,z')\right)\Qloc(z')\\ \\
  &&\qquad\qquad-\sin\left(2\bt(z_1,z')\right)\Uloc(z')\Big)dz';\\ \\
  U&=&\int_{z_1}^{z_2}\Big(\sin\left(2\bt(z_1,z')\right)\Qloc(z')\\ \\
  &&\qquad\qquad+\cos\left(2\bt(z_1,z')\right)\Uloc(z')\Big)dz'.
\end{array}
\right.
\end{equation}
The polarization fraction can be then derived as
\begin{equation}
\Pi=\frac{\sqrt{Q^2+U^2}}{I}.
\end{equation}
The angle
\begin{equation}
\Psi=\frac{1}{2}\arctan\left(\frac{U}{Q}\right).
\end{equation}
is orthogonal to the polarization angle, and gives the average orientation of the ``observed'' MF, namely we have
\begin{equation}
B_x=B\cos\Psi,\qquad B_y=B\sin\Psi. 
\end{equation}

\subsection{Single-layer case}

The problem simplifies considerably within the thin-layer approximation, namely under the assumption that $n\Bpar$, as well as the specific $\Qloc$ and $\Uloc$ values, are constant across the shell.
In this case, since
\begin{equation}
\frac{d\bt}{dz'}=\frac{e^3\lmb^2}{2\pi\me^2c^4}n\Bpar
\end{equation}
is constant, the integration along a layer ($z_1=0$, $z_2=z$) leads to the following relations:
\begin{eqnarray}
\label{eq:intgsinbt}
\int_0^z{\sin\left(2\frac{d\bt}{dz'}z'\right)}dz'&=&z\frac{\sin\bt}{\bt}\sin\bt;	\\ 
\label{eq:intgcosbt}
\int_0^z{\cos\left(2\frac{d\bt}{dz'}z'\right)}dz'&=&z\frac{\sin\bt}{\bt}\cos\bt.
\end{eqnarray}
In addition, we readily have:
\begin{equation}
z\Qloc=\Qi;\qquad z\Uloc=\Ui.
\end{equation}
Then combining all together, we finally get
\begin{eqnarray}
\label{eq:Qsingle}
Q&=&\frac{\sin(\bt)}{\bt}\Big(\cos(\bt)\Qi-\sin(\bt)\Ui\Big);	\\ 
\label{eq:Usingle}
U&=&\frac{\sin(\bt)}{\bt}\Big(\sin(\bt)\Qi+\cos(\bt)\Ui\Big).
\end{eqnarray}
Namely, the amount of rotation is half of that for a foreground layer; while, differently from the case of a foreground FR, which turns out into a mere rotation of the polarization plane, now the observed polarization fraction is lower than the intrinsic one.
In fact
\begin{equation}
\sqrt{Q^2+U^2}=\left|\frac{\sin(\bt)}{\bt}\right|\sqrt{\Qi^2+\Ui^2}.
\end{equation}

\subsection{Two-layers combination}
In the case of a thin shell SNR, in each projected position we see the combination of the effects of two layers, each of them subject to internal FR, and in addition the front-side layer Faraday rotating the polarized emission from the rear-side one (in the following we will label the front and rear limb quantities as ``f'' and ``r'' respectively).
We then have
\begin{eqnarray}
Q\!\!&=&\!\!\frac{\sin(\RMf\lmb^2)}{\RMf\lmb^2}\Big(\cos(\RMf\lmb^2)\Qf-\sin(\RMf\lmb^2)\Uf\Big)
	\nonumber	\\
&&\!\!\!\!\!\!\!\!\!\!\!\!\!\!\!\!\!\!
+\frac{\sin(\RMr\lmb^2)}{\RMr\lmb^2}\Big(\cos(\RMx\lmb^2)\Qr-\sin(\RMx\lmb^2)\Ur\Big);	\\ 
U\!\!&=&\!\!\frac{\sin(\RMf\lmb^2)}{\RMf\lmb^2}\Big(\sin(\RMf\lmb^2)\Qf+\cos(\RMf\lmb^2)\Uf\Big)
	\nonumber	\\
&&\!\!\!\!\!\!\!\!\!\!\!\!\!\!\!\!\!\!
+\frac{\sin(\RMr\lmb^2)}{\RMr\lmb^2}\Big(\sin(\RMx\lmb^2)\Qr+\cos(\RMx\lmb^2)\Ur\Big),
\end{eqnarray}
where $\RMx=2\RMf+\RMr$.

Even in the simplified case of two homogeneous thin layers, the formulae for $\fobs$ and $\RMobs$ (as defined by Eqs.~\ref{eq:observedQFR} and \ref{eq:observedQFR}) turn out to be rather complex.
For this reason we consider here the limit for small rotations, namely for small enough $\lmb$ values, getting
\begin{eqnarray}
\label{eq:fobsthin}
\fobs\!\!\!\!\!&=&\!\!\!\!\!1-\frac{\QUm\RMt}{(\Qi^2+\Ui^2)}\lmb^2+\Ord(\lmb^4);		\nonumber\\ 
&=&\!\!\!\!\!1-\fobstwo\lmb^2+\Ord(\lmb^4);		\\
\RMobs\!\!\!\!\!&=&\!\!\!\!\!\left(\frac{\RMf}{2}+\frac{\QUp\RMt)}{2(\Qi^2+\Ui^2)}\right)	\nonumber\\
&&\!\!\!\!\!\!\!\!\!\!\!\!\!\!\!\!\!\!\!\!\!
+\QUm\left(\frac{\RMt\RMx}{6(\Qi^2+\Ui^2)}-\frac{\QUp\RMt^2}{2(\Qi^2+\Ui^2)^2}\right)\lmb^2+\Ord(\lmb^4)														\nonumber\\
\label{eq:RMobsthin}
&=&\!\!\!\!\!\RMobszero+\RMobstwo\lmb^2+\Ord(\lmb^4),
\end{eqnarray}
where $\QUm=\Qf\Ur-\Qr\Uf$, $\QUp=\Qi\Qr+\Ui\Ur$, $\RMt=\RMf+\RMr$, $\Qi=\Qf+\Qr$ and $\Ui=\Uf+\Ur$.
Note that the quantity $\fobs$, previously dubbed as ``depolarization factor'' does not always lead to a decreased polarization level.
In its series expansion, it depends on the sign of the coefficient of $\lmb^2$).
This means that under some circumstances the differential FR may have a constructive effect.

At least in principle, comparing polarizations measured at various wavelengths, including longer ones, would allow one to measure more coefficients of the power expansions with $\lmb$ of both $\RMobs$ and $\fobs$, then allowing to get further information on the MF structure.
This is unfortunately a difficult observational task at the present time, but it should become feasible with the advent of larger radiotelescopes operating at low-frequency radio wavelengths (like in the cases of LOFAR and SKA).

\subsection{A generalized momentum treatment}

Releasing the thin-layer assumption, the treatment of the internal FR becomes considerably more complex, since the calculation of the observed Stokes parameters requires the explicit calculation of a double integral for each projected point.
A complete solution of the inverse problem, namely how to derive from the observations the 3-D structure of density and MF, looks then essentially hopeless.
However, in an analogous way to that shown in the previous section, some relation can be derived as power-law expansions, in the limit of wavelengths.

The observed Stokes parameters read
\begin{eqnarray}
Q\!\!\!\!\!\!&=&\!\!\!\!\!\!\!\!\int{\!\!\cos\!\Big(\!2\RM(z)\lmb^2\!\Big)\Qloc(z)-\sin\!\Big(\!2\RM(z)\lmb^2\!\Big)\Uloc(z)\,dz};	\\ 
U\!\!\!\!\!\!&=&\!\!\!\!\!\!\!\!\int{\!\!\sin\!\Big(\!2\RM(z)\lmb^2\!\Big)\Qloc(z)+\cos\!\Big(\!2\RM(z)\lmb^2\!\Big)\Uloc(z)\,dz}.
\end{eqnarray}
Let us introduce the quantities
\begin{eqnarray}
\Qin{n}&=&\int_0^\infty{\Qloc(z)\RM(z)^n}\,dz;	\\
\Uin{n}&=&\int_0^\infty{\Uloc(z)\RM(z)^n}\,dz
\end{eqnarray}
(in the following, for $\Qin{0}$ and $\Uin{0}$ we shall simply use $\Qi$ and $\Ui$ respectively).
It can be shown that the power expansions of both observed depolarization $\fobs$ and rotation measure $\RMobs$ can be expressed as fractions whose numerator is a polynomial of these $I$ quantities, while the denominator is a power of $(\Qi^2+\Ui^2)$.
We present here the expansions equivalent to those shown by Eqs.~\ref{eq:fobsthin} and \ref{eq:RMobsthin} for the thin-layer case
\begin{eqnarray}
\label{eq:fobsgen}
\fobs\!\!\!\!\!&=&\!\!\!\!\!1-\frac{2(\Qi\Uin{1}-\Ui\Qin{1})}{(\Qi^2+\Ui^2)}\lmb^2+\Ord(\lmb^4);	\\ 
\label{eq:RMobsgen}
\RMobs\!\!\!\!\!&=&\!\!\!\!\!\frac{\Qi\Qin{1}+\Ui\Uin{1}}{(\Qi^2+\Ui^2)}	
+\left(\frac{\Ui\Qin{2}-\Qi\Uin{2}}{(\Qi^2+\Ui^2)}+\right.
\nonumber\\
&&
\!\!\!\!\!\!\!\!\!\!\!\!\!\!\!\!\!\!\!\!\!\!\!\!\!\!\!\!\!\!\!\!
\left.+\frac{2\left((\Qi^2-\Ui^2)\Qin{1}\Uin{1}-\Qi\Ui(\Qin{1}^2-\Uin{1}^2)\right)}{(\Qi^2+\Ui^2)^2}\!\!\right)\!\!\lmb^2\!\!+\Ord(\lmb^4).
\end{eqnarray}
Eqs.~\ref{eq:fobsthin} and \ref{eq:RMobsthin} are indeed recovered, by substituting
\begin{eqnarray}
\Qin{1}\!\!\!\!&=&\!\!\!\!\Qr\RMf+\frac{\Qf\RMf+\Qr\RMr}{2};	\\
\Uin{1}\!\!\!\!&=&\!\!\!\!\Ur\RMf+\frac{\Uf\RMf+\Ur\RMr}{2};	\\
\Qin{2}\!\!\!\!&=&\!\!\!\!\Qr\RMf^2+\frac{\Qr\RMf\RMr}{2}
+\frac{\Qf\RMf^2+\Qr\RMr^2}{3};	\\
\Uin{2}\!\!\!\!&=&\!\!\!\!\Ur\RMf^2+\frac{\Ur\RMf\RMr}{2}
+\frac{\Uf\RMf^2+\Ur\RMr^2}{3}
\end{eqnarray}
Here is a list of general conclusions arising from this analysis (and that in some respects, confirm what already found from the thin-layer approximated analysis):\\
- Differently from what happens when the FR is just due to foreground medium, the internal FR does not simply scale with $\lmb^2$.\\
- While foreground FR does not cause depolarization (apart from that produced by the finite spectral width of the telescope), internal FR necessarily involves some change in the polarization level, which depending on the case may lead to a depolarization or to an increase of the polarization fraction.\\
- The momentum analysis can in principle provide an infinite number of observational constraints: they correspond to the various orders of the power expansion in $\lmb$ of Eqs.~\ref{eq:fobsgen} and \ref{eq:RMobsgen} that, combined to the fact that $Q\rightarrow\Qi$ and $U\rightarrow\Ui$, for $\lmb\rightarrow0$.\\
- Unfortunately, the number of relations is not sufficient to derive $\Qloc(z)$, $\Uloc(z)$, and $\RM(z)$ independently (but just two of them, given the third one): therefore some a priori conditions on the MF and the emissivity must be given.

\begin{figure*}
\centering
\includegraphics[angle=0,width=16.0truecm]{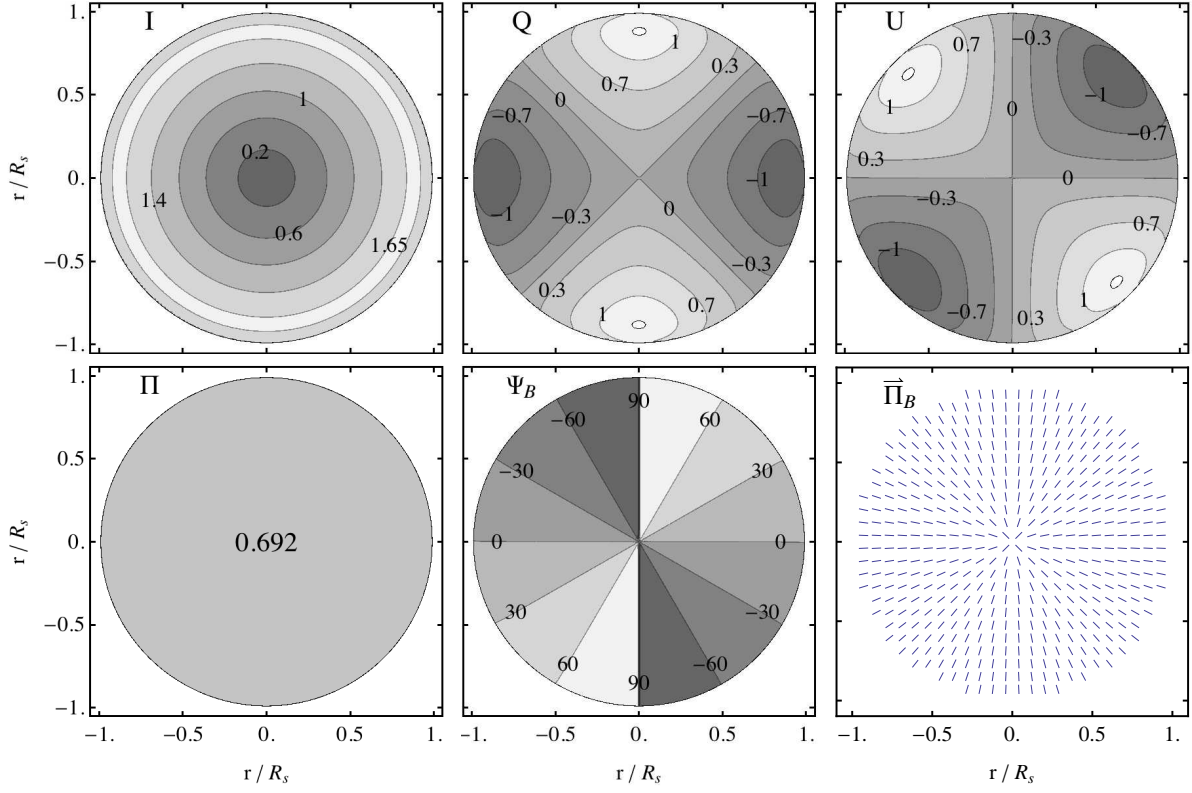}
\caption{Case of fully ordered, meridionally oriented MF, without FR: aspect angle of $0^\circ$. See text for the explanation of the panels.}
\label{fig:meridFOnoFRang0}
\end{figure*}
\begin{figure*}
\centering
\includegraphics[angle=0,width=16.0truecm]{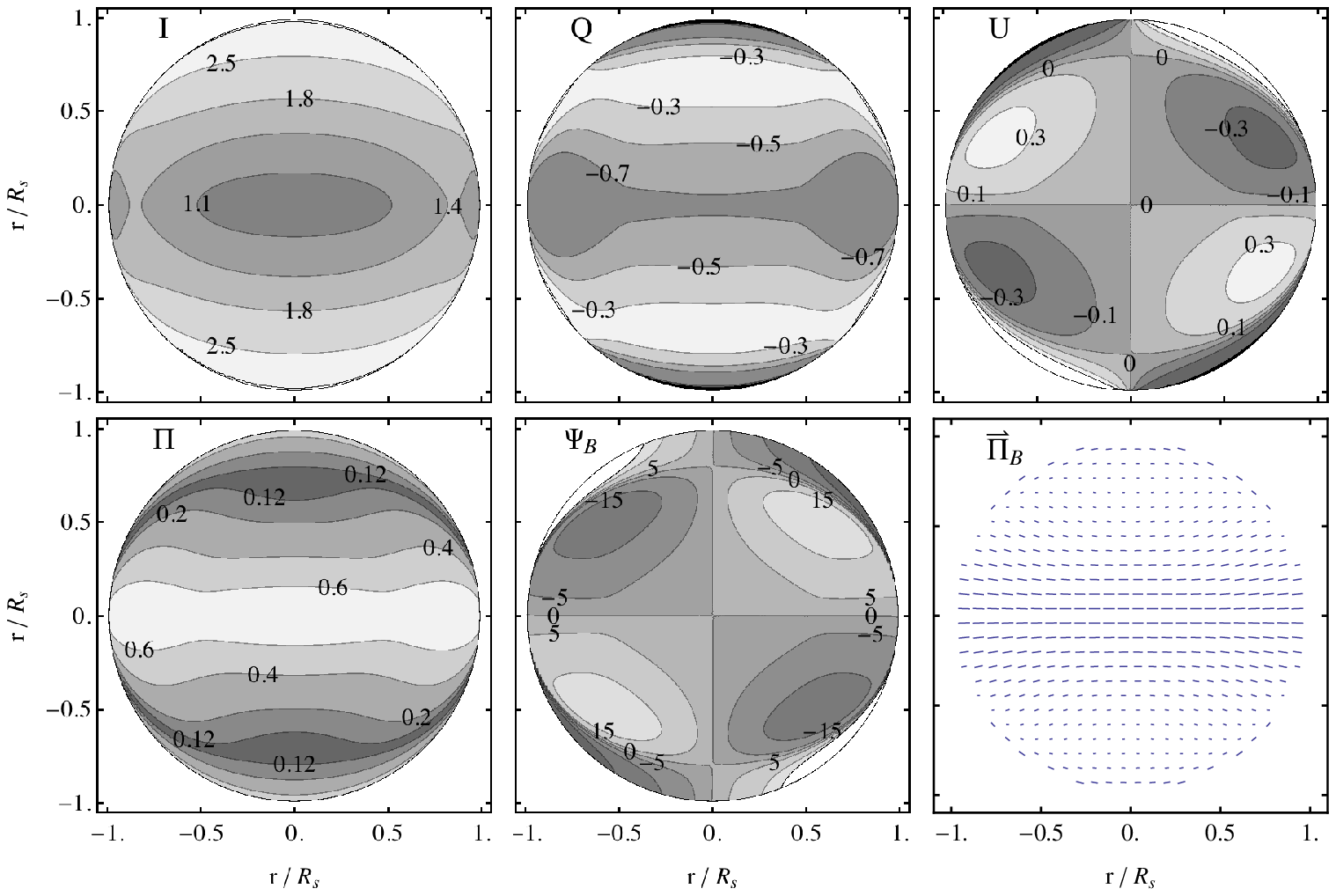}
\caption{Case of fully ordered, meridionally oriented MF, without FR: aspect angle of $30^\circ$. }
\label{fig:meridFOnoFRang30}
\end{figure*}
\begin{figure*}
\centering
\includegraphics[angle=0,width=16.0truecm]{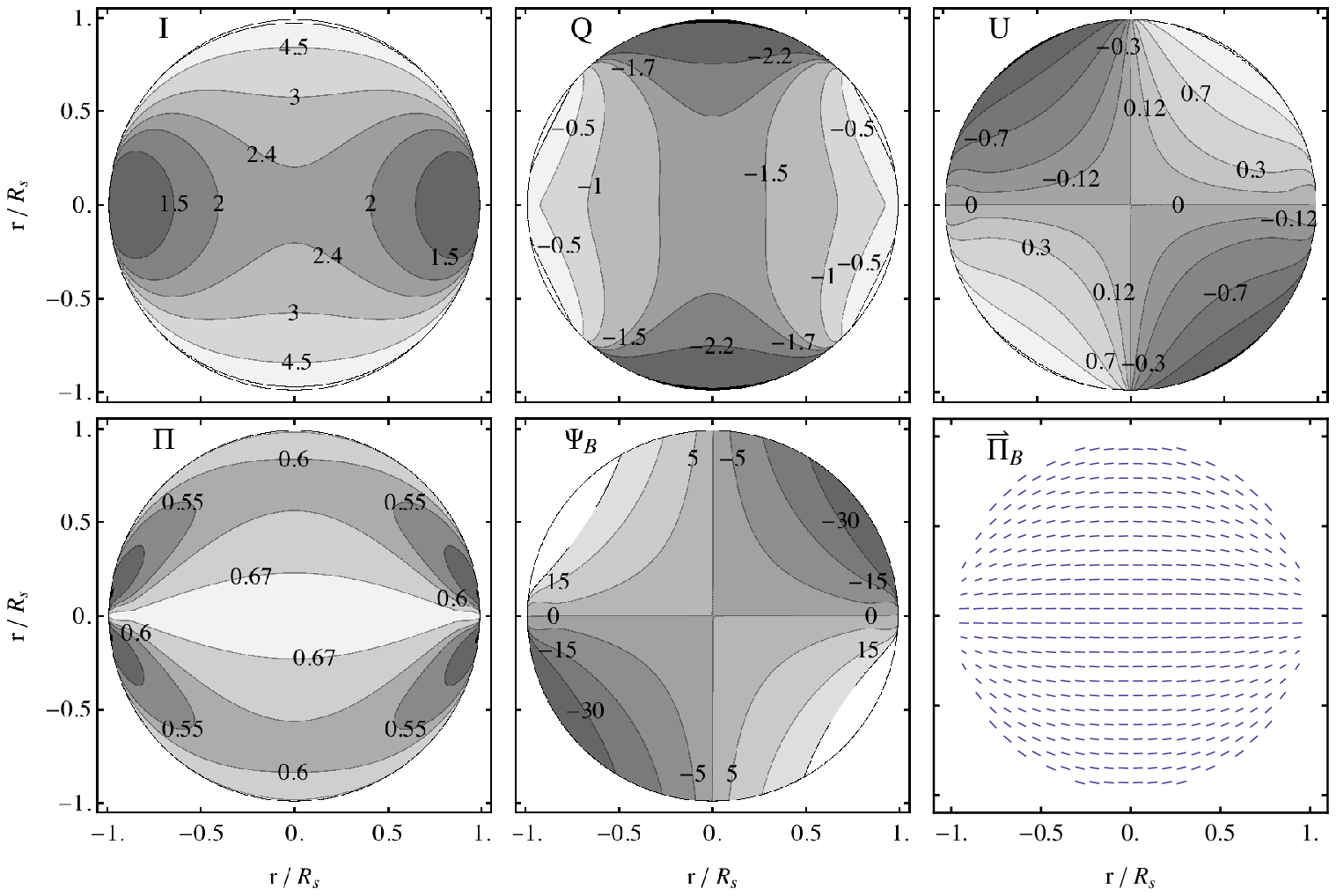}
\caption{Case of fully ordered, meridionally oriented MF, without FR: aspect angle of $60^\circ$. }
\label{fig:meridFOnoFRang60}
\end{figure*}
\begin{figure*}
\centering
\includegraphics[angle=0,width=16.0truecm]{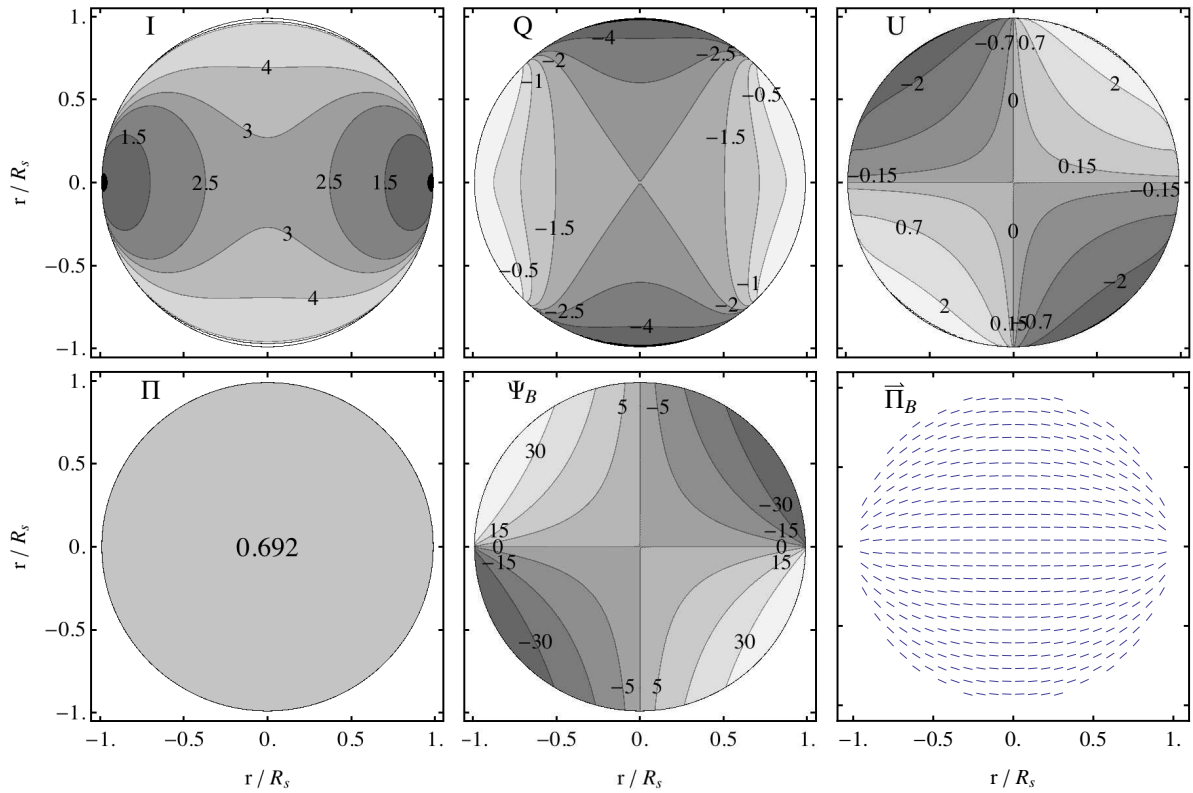}
\caption{Case of fully ordered, meridionally oriented MF, without FR: aspect angle of $90^\circ$. }
\label{fig:meridFOnoFRang90}
\end{figure*}

\section{Thin-layer models of SNRs}

In this section we present and discuss some SNR models generated using the thin-layer approach described above.
It is not our intention, at this stage, to model any individual SNR, also because real maps typically show patterns at smaller scales, which will not be considered here.
Our present aim is just that of outlining some general trends, for a better insight on the importance of the various parameters.
For all models presented below we have used a shell thickness $10^{-2}$ times its radius , and a power-law particle energy distribution with $s=2$ (corresponding to a radio spectral index of 0.5, close to what usually measured in SNRs).
The exact value of the thickness ($w$, here taken to be constant), provided it is much smaller than $\Rs$, does not affect considerably the pattern of the emission maps, but just their flux normalization.

It must be noticed that all maps of surface brightness presented below are normalized to
\begin{equation}
\Wz\Bz^{(s+1)/2}w,
\end{equation}
where the quantities $\Wz$ and $\Bz$ are their respective maximum values within the shell: for instance, the equatorial value (namely that at $\tht=\pi/2$), for a meridional MF like that defined above.
In the following maps (except when stated differently) the projection of the SNR symmetry axis runs from left to right; while for all maps in tones of grey, lighter tones means higher values.
\begin{figure*}
\centering
\includegraphics[angle=0,width=16.0truecm]{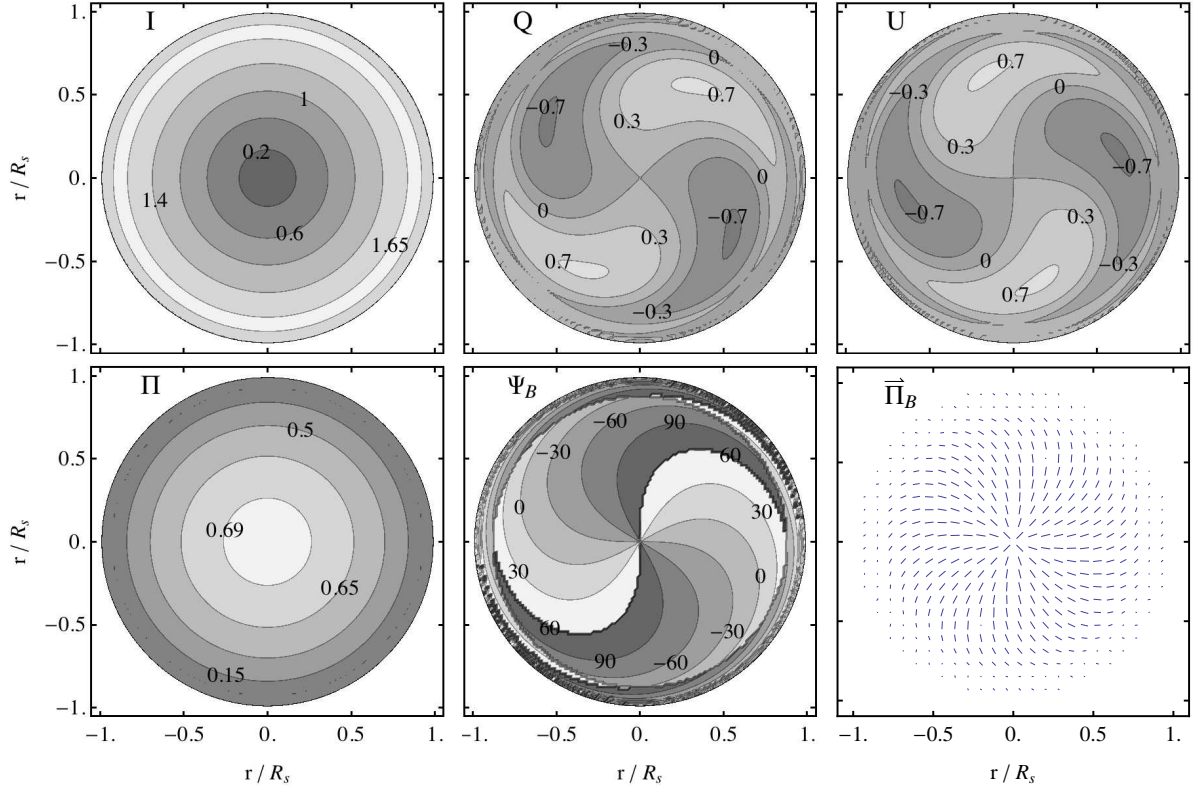}
\caption{Case of fully ordered, meridionally oriented MF, with FR: aspect angle of $0^\circ$.}
\label{fig:meridFOFRang0}
\end{figure*}
\begin{figure*}
\centering
\includegraphics[angle=0,width=16.0truecm]{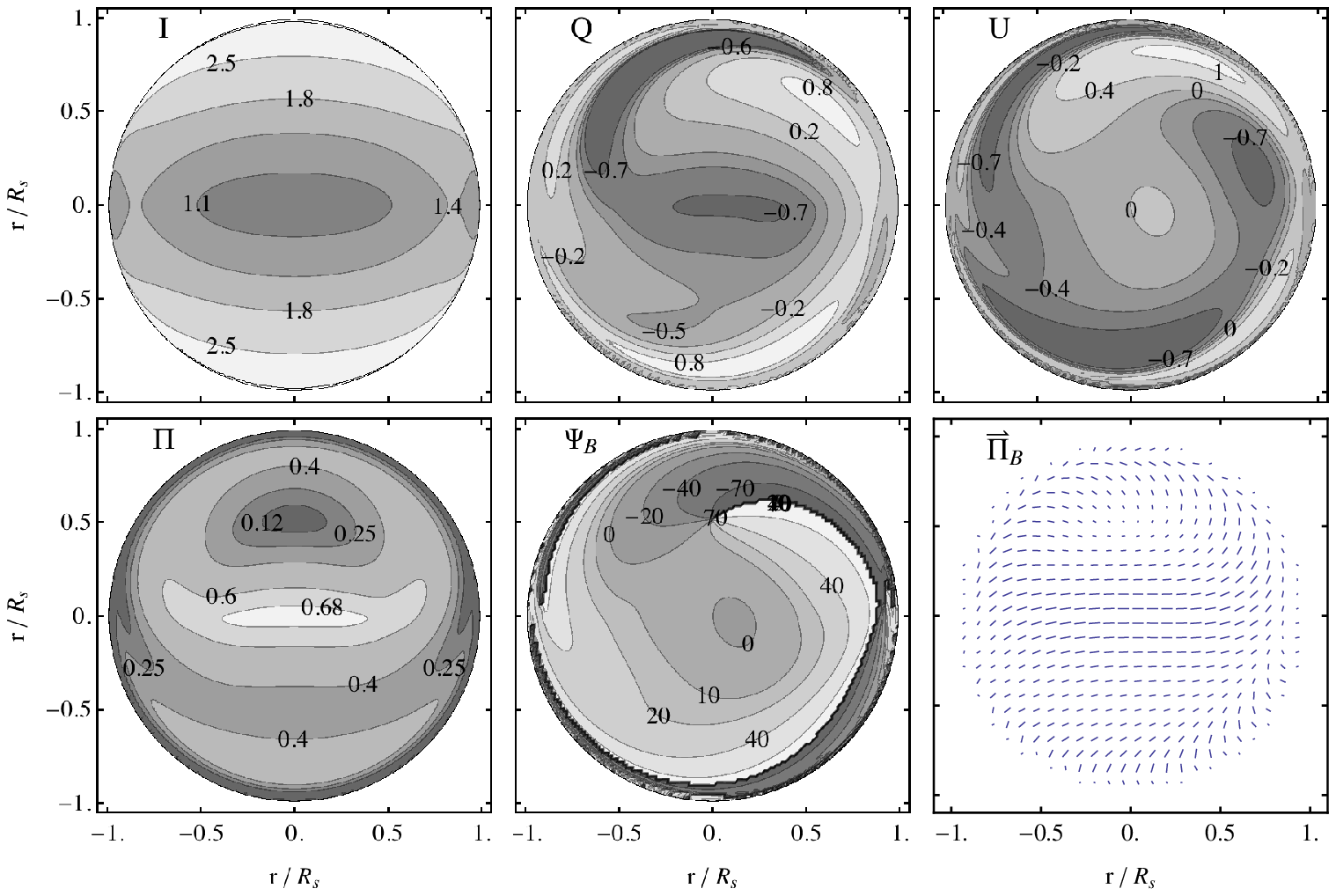}
\caption{Case of fully ordered, meridionally oriented MF, with FR: aspect angle of $30^\circ$. }
\label{fig:meridFOFRang30}
\end{figure*}
\begin{figure*}
\centering
\includegraphics[angle=0,width=16.0truecm]{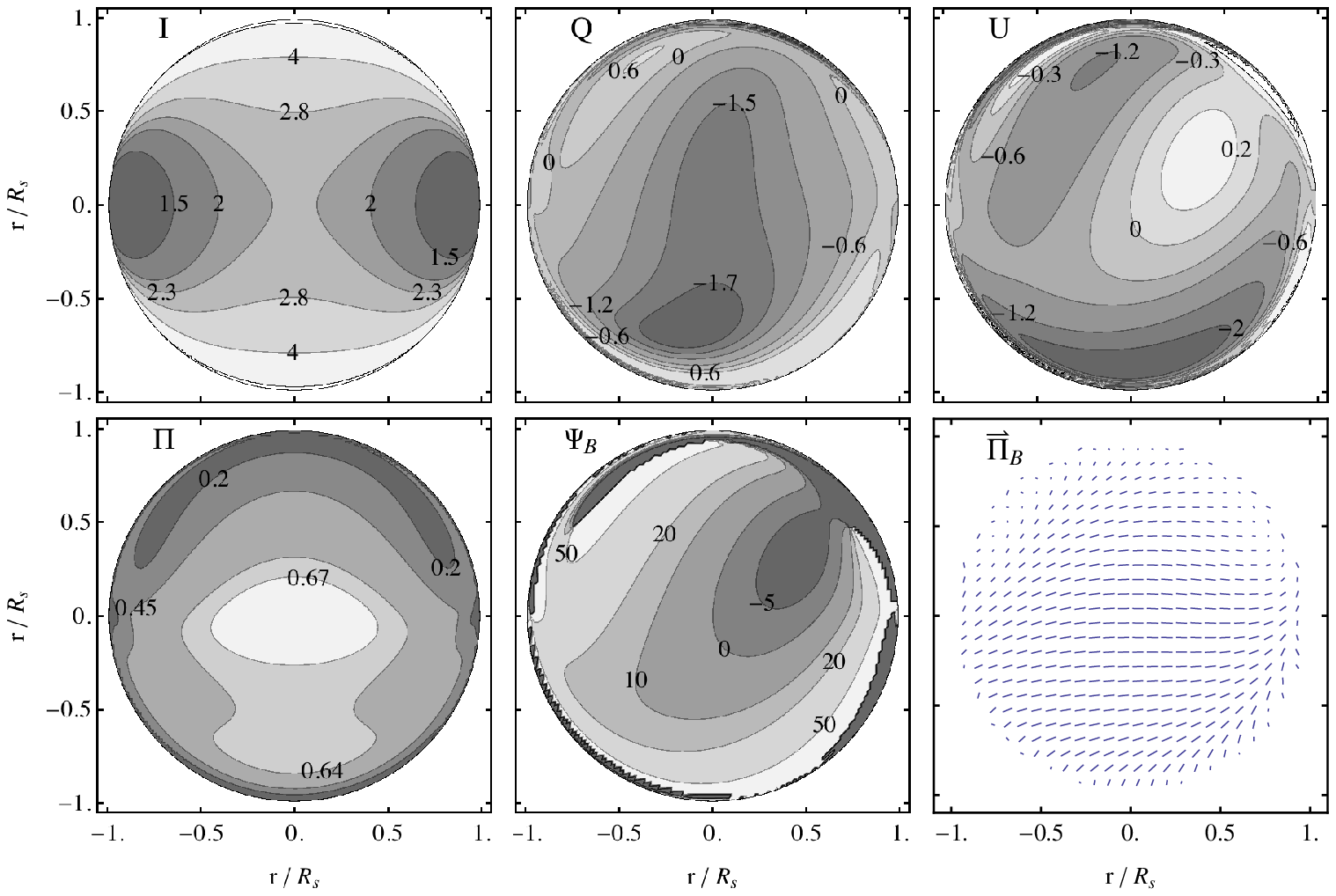}
\caption{Case of fully ordered, meridionally oriented MF, with FR: aspect angle of $60^\circ$. }
\label{fig:meridFOFRang60}
\end{figure*}
\begin{figure*}
\centering
\includegraphics[angle=0,width=16.0truecm]{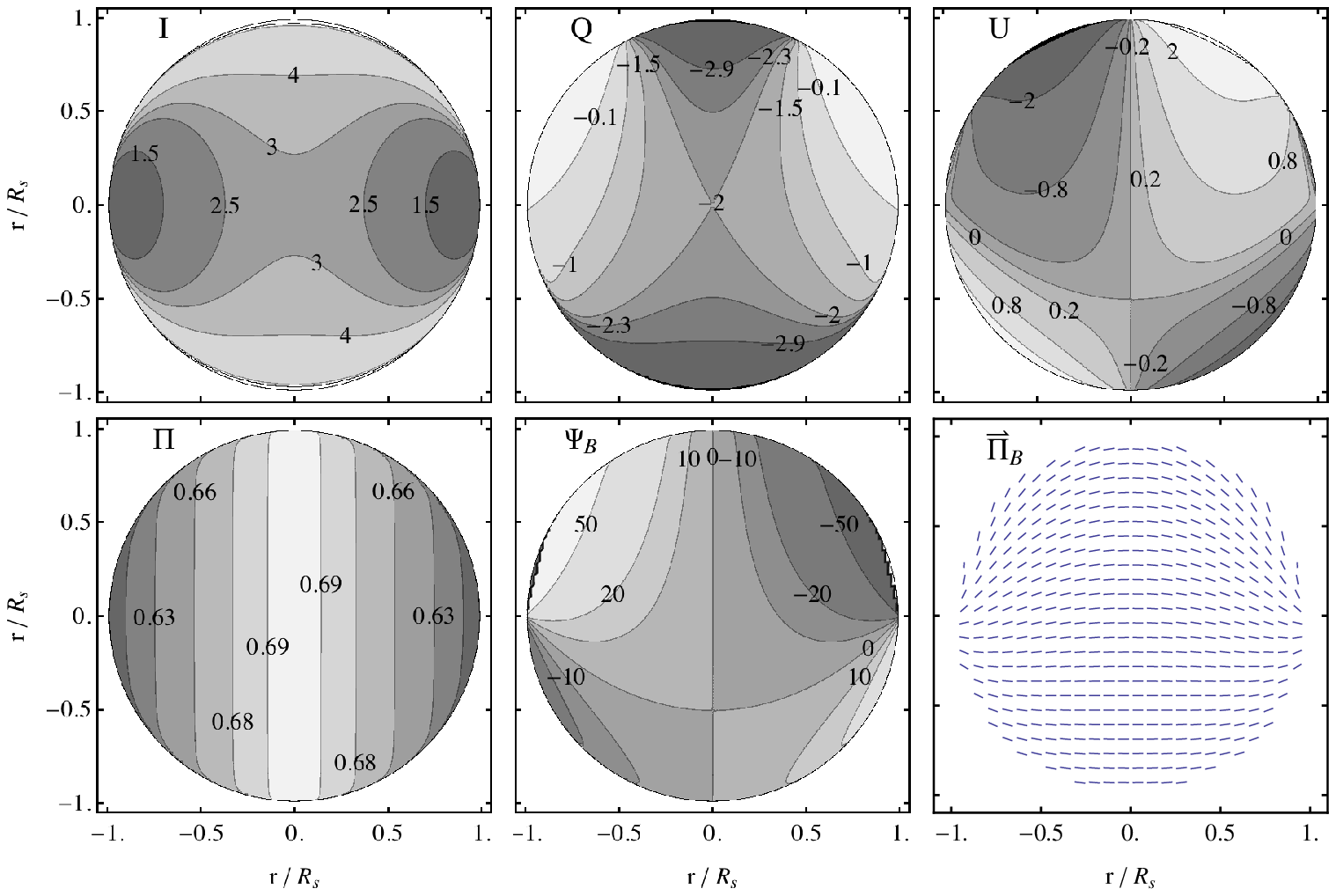}
\caption{Case of fully ordered, meridionally oriented MF, with FR: aspect angle of $90^\circ$. }
\label{fig:meridFOFRang90}
\end{figure*}

\subsection{Fully ordered MF and negligible FR}

Let us begin considering the case of a completely ordered MF, aligned along the meridians, as from Eqs.~\ref{eq:uvBmeridional}--\ref{eq:magBmeridional} with constant layer thickness.
Let us also assume that the normalization $A$ of the particle distribution is a constant (different functional dependencies with $\tht$ will also be used, then changing the relative importance of the emissivity, for instance, of the equatorial zone with respect to the polar ones).

At this stage we take an observing frequency high enough that FR does not give any appreciable effect.
The results are shown in Fig.~\ref{fig:meridFOnoFRang0}, \ref{fig:meridFOnoFRang30}, \ref{fig:meridFOnoFRang60}, and \ref{fig:meridFOnoFRang90}, respectively for aspect angles of  $0^\circ$ (the polar direction),  $30^\circ$,  $60^\circ$, and $90^\circ$ (along the equatorial plane): this choice of aspect angles allows one to appreciate the continuity of the change of properties between the limit cases, even though they look so different one from the other.

The three upper panels, from left to right, show the maps in the $I$, $Q$ and $U$ Stokes parameters; while the three lower panels show more intuitive polarization quantities namely, again from left to right, the  polarization fraction, the magnetic polarization angle with respect to the projected axis, and finally the vectorial map of the magnetic polarization.

In the $0^\circ$ and $90^\circ$ special cases the direction of projected MF in the front layer coincides with that in the rear layer: therefore, in the absence of FR, the measured polarization is spatially constant, and  reaches its maximum theoretical value (for all these maps we have assumed $s=2$, equivalent to a spectral index of --0.5 for the emission).

\subsection{The effect of FR}

In the same way as we did with the emission, let us use dimensionless quantities also for the treatment of FR.
With reference to Eq.~\ref{eq:polbetadef}, let us normalize all RM quantities with:
\begin{equation}
\frac{e^3}{2\pi\me^2c^4}\nz\Bz w
\end{equation}
where $\nz$ is the electron density in the shell.
In Fig.~\ref{fig:meridFOFRang0}, \ref{fig:meridFOFRang30}, \ref{fig:meridFOFRang60}, and \ref{fig:meridFOFRang90} we may see the kind of distortion produced by FR: here we have used an intermediate value of RM, namely such to give a FR equal to unity at the equator, when seen face-on.
The results are shown in Fig.~\ref{fig:meridFOFRang0}, \ref{fig:meridFOFRang30}, \ref{fig:meridFOFRang60}, and \ref{fig:meridFOFRang90}, respectively for the same aspect angles of $0^\circ$, $30^\circ$, $60^\circ$, and $90^\circ$: also in this case, even though the patterns are distorted, one may appreciate the continuity in the change of pattern between contiguous aspect angles.

\begin{figure*}
\centering
\includegraphics[angle=0,width=16.0truecm]{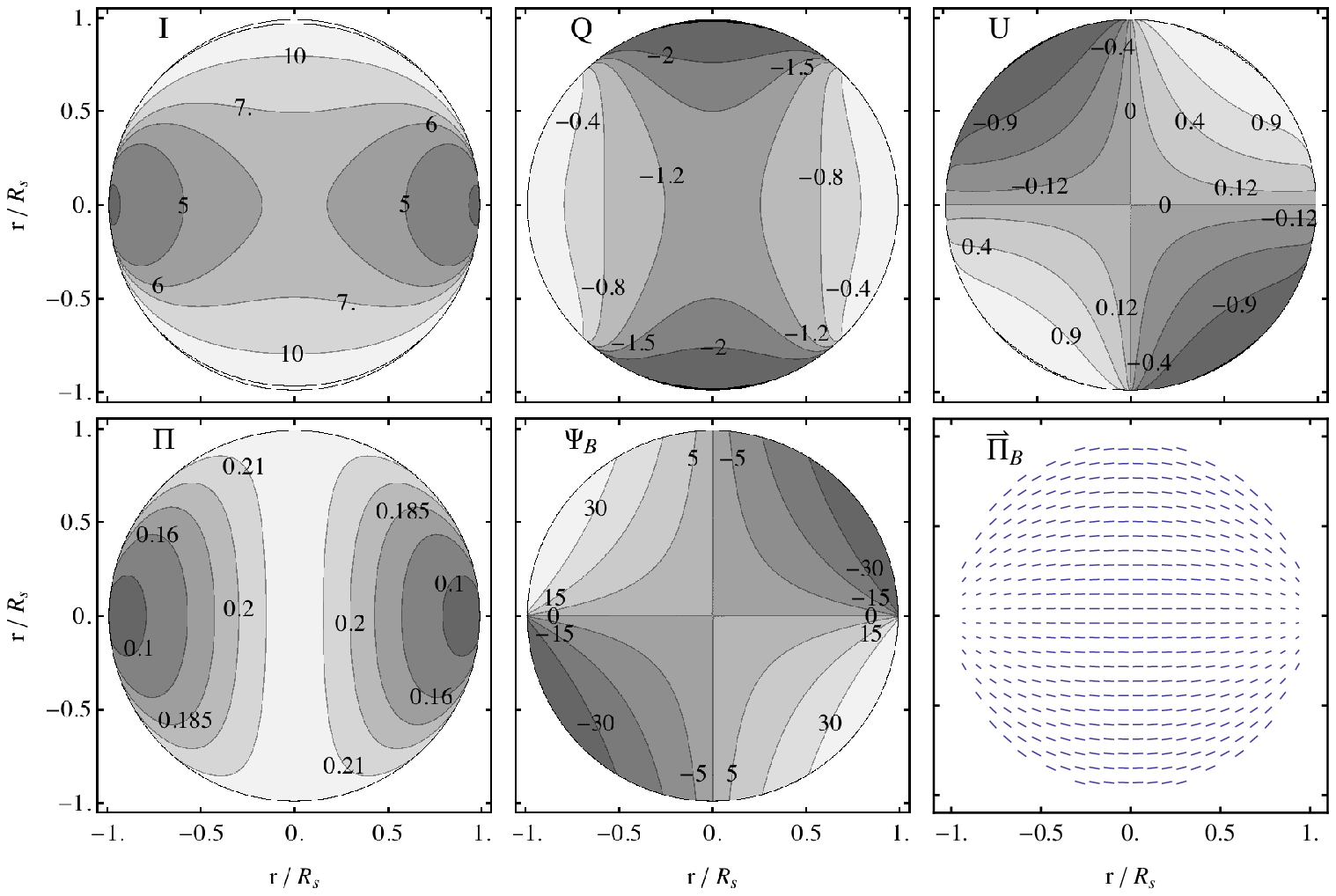}
\caption{Fully ordered, meridionally oriented MF, with $\sg/\Bav=1$: aspect angle of $90^\circ$. }
\label{fig:meridFOSGang90}
\end{figure*}
\subsection{The effect of a random MF component}

Let us consider here just the case in which in addition to the ordered MF there is an isotropic random component, in which case the emission quantities derive from Eqs.~\ref{eq:IPL} and \ref{eq:QPL}.
While there is an infinity of possible cases, here we shall just consider the simplest one (even though not physically more justified than others), in which the $\sg/\Bav$ ratio is constant everywhere; moreover, with the aim of discussing the various effects independently, let us consider here just the case with vanishing FR.
For an isotropic random MF component, a constant $\sg/\Bav$ ratio means that on the projected MF the effect of fluctuations will be larger when the ordered MF is oriented closer to the line of sight, and therefore the depolarization level will be higer in those cases.
The effect is more evident at large aspect angles, so that in Fig.~\ref{fig:meridFOSGang90} we present the case at an aspect angle of $90^\circ$ (and $\sg/\Bav=1$): this map has to be compared with Fig.~\ref{fig:meridFOnoFRang90}.

\begin{figure*}
\centering
\includegraphics[angle=0,width=16.0truecm]{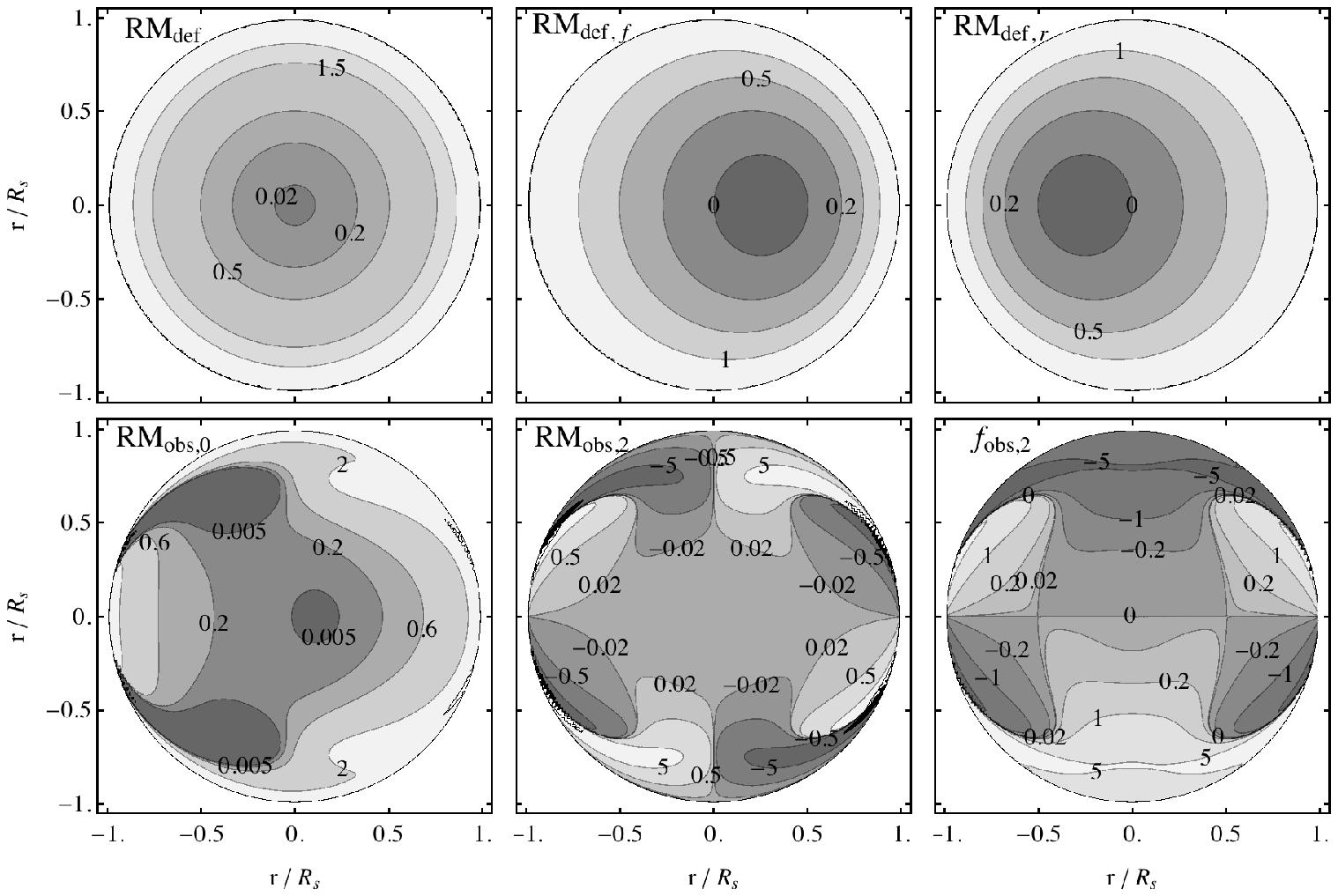}
\caption{FR-related quantities, for the case shown in Fig.~\ref{fig:meridFOFRang30} (aspect angle of $30^\circ$).}
\label{fig:meridRMFRang30}
\end{figure*}
\begin{figure*}
\centering
\includegraphics[angle=0,width=16.0truecm]{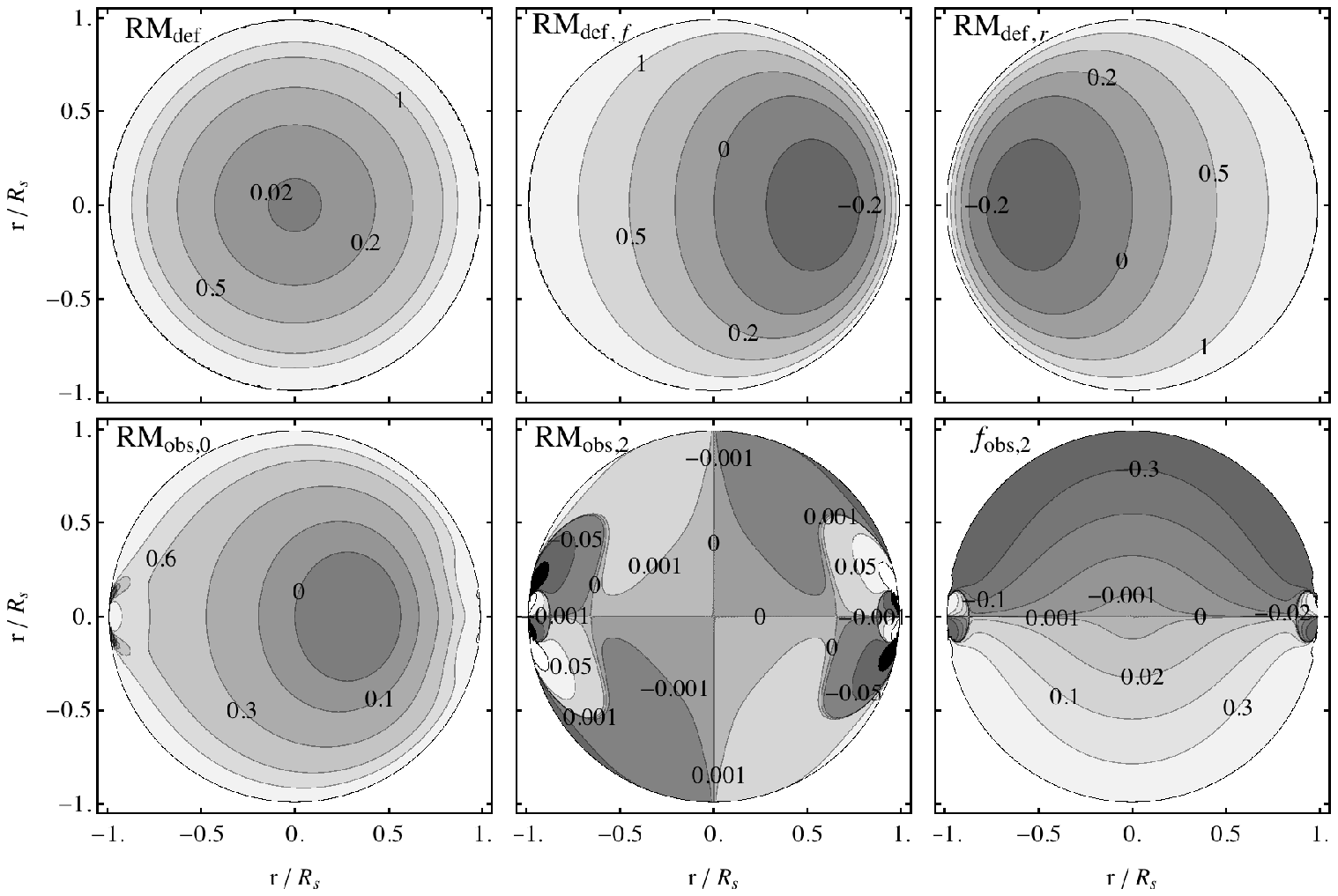}
\caption{FR-related quantities, for the case shown in Fig.~\ref{fig:meridFOFRang60} (aspect angle of $60^\circ$).}
\label{fig:meridRMFRang60}
\end{figure*}
\begin{figure*}
\centering
\includegraphics[angle=0,width=16.0truecm]{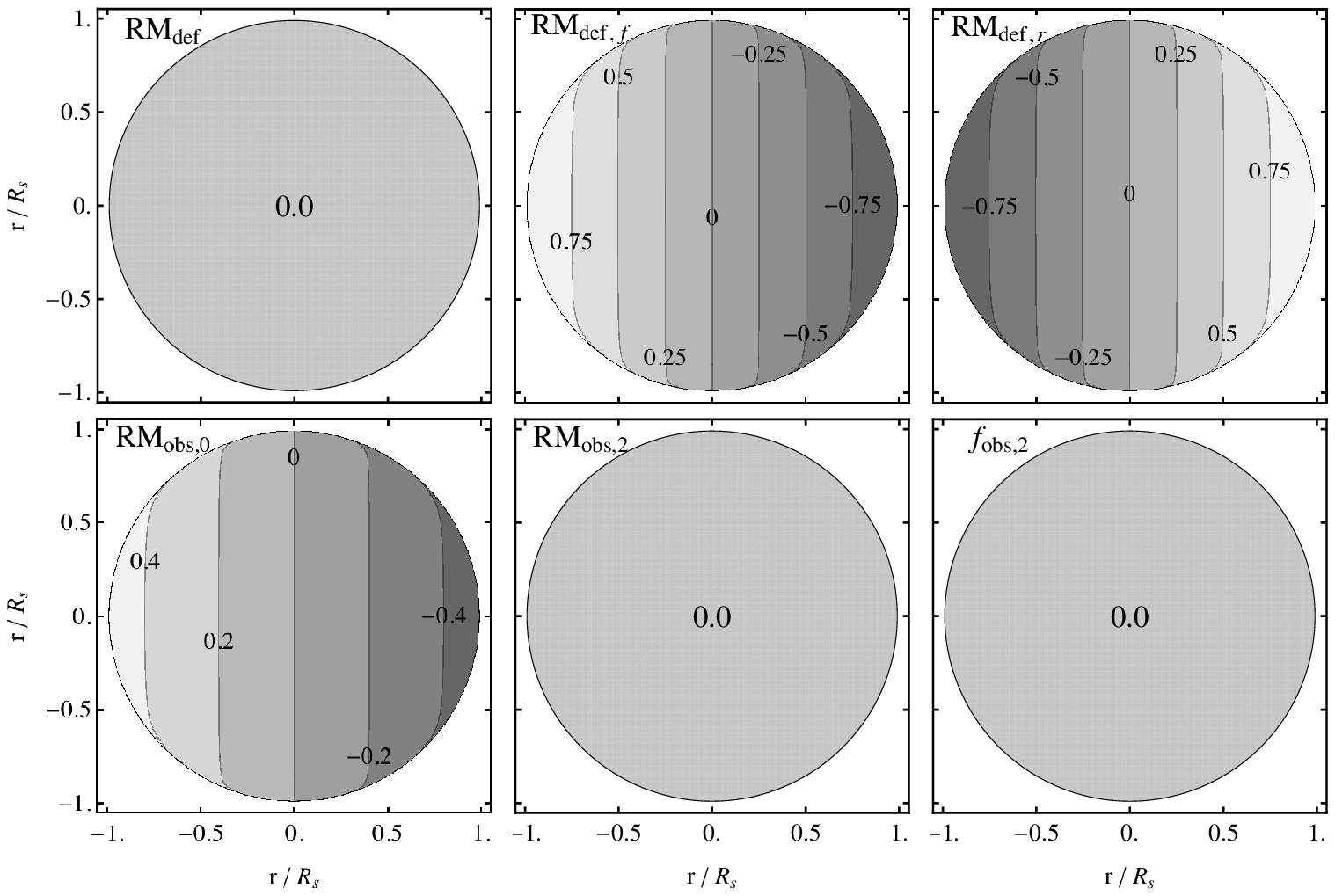}
\caption{FR-related quantities, for the case shown in Fig.~\ref{fig:meridFOFRang90} (aspect angle of $90^\circ$).}
\label{fig:meridRMFRang90}
\end{figure*}
\subsection{Structure of RM maps}

In the previous sections we have already made a distinction between the standard definition of the RM (Eq.~\ref{eq:polarizRMdef}, which can be associated to real observations only in the case of FR of a background source), and the RM (plus depolarization) that can be actually derived from observations in the case of intrinsic FR (see Eqs.~\ref{eq:fobsthin} and \ref{eq:RMobsthin}).
Figs.~\ref{fig:meridRMFRang30} and \ref{fig:meridRMFRang90}, respectively associated to Figs.~\ref{fig:meridFOFRang30} and \ref{fig:meridFOFRang90}, show these FR-related quantities: in the upper panels the RM as from the standard definition, namely (from left to right) its total value, the value for the front layer, and that for the rear layer; in the lower panels, instead, (again from left to right) the observed map of the intrinsic RM (more precisely $\RMobszero$, its small $\lmb$ limit), and then the quantities tracing the next order of approximation effects respectively on the polarization direction ($\RMobstwo$; the coefficient of $\lmb^2$, in Eq.~\ref{eq:RMobsthin}) and on the polarization degree ($\fobstwo$; the coefficient of $\lmb^2$, with positive sign, in Eq.~\ref{eq:fobsthin}).

The next order of approximation effects, which to our knowledge have not been measured as yet, should be observable with high-quality and high-resolution observations at low-frequency radio wavelength: testing their patterns would provide a further information to constrain the structure of a radio SNR.
We may see that, at an aspect angle of $90^\circ$, the second-order terms vanish: in general, one may expect them to vanish all the times in which the intrinsic polarization angles of the front and rear layer are equal, this implying the quantity $\QUm$ (defined right after Eq.~\ref{eq:RMobsthin}) to vanish.

\begin{figure*}
\centering
\includegraphics[angle=0,width=16.0truecm]{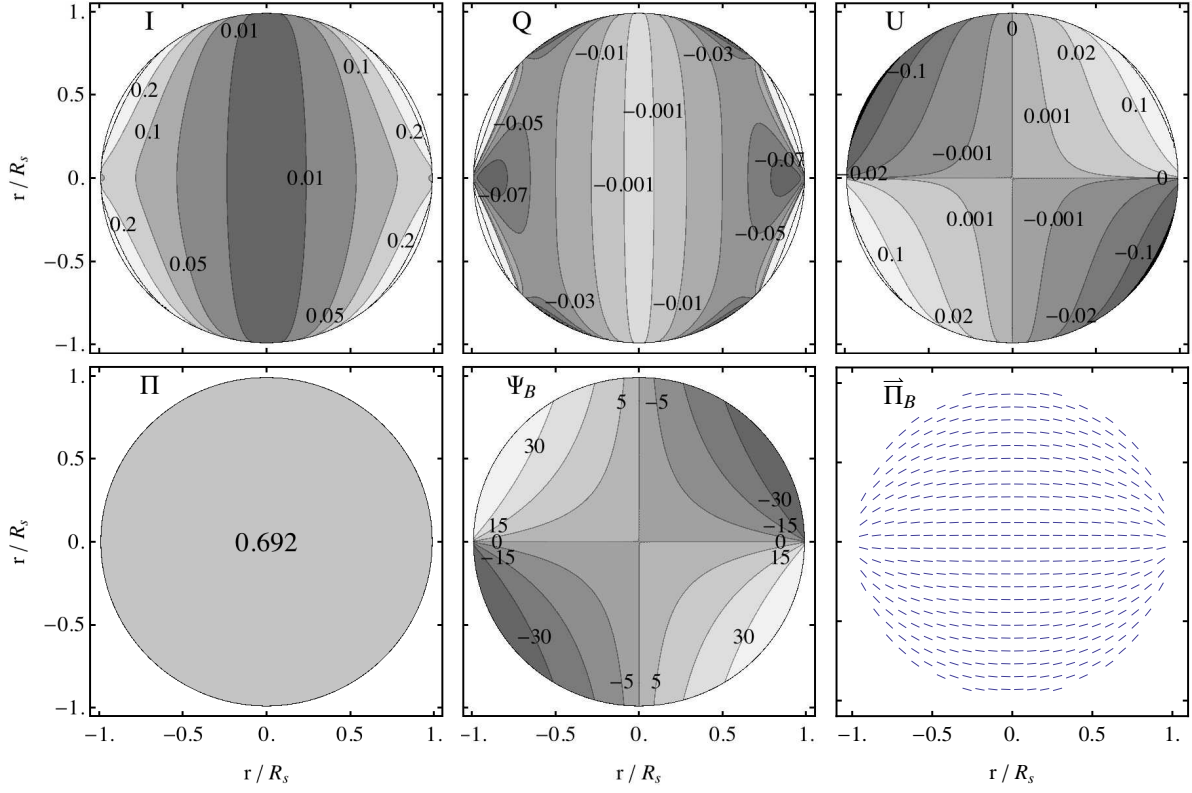}
\caption{Case of a meridionally oriented MF, with quasi-parallel injection efficiency: aspect angle of $90^\circ$. }
\label{fig:meridFOnoFRang90PARA}
\end{figure*}
\begin{figure*}
\centering
\includegraphics[angle=0,width=16.0truecm]{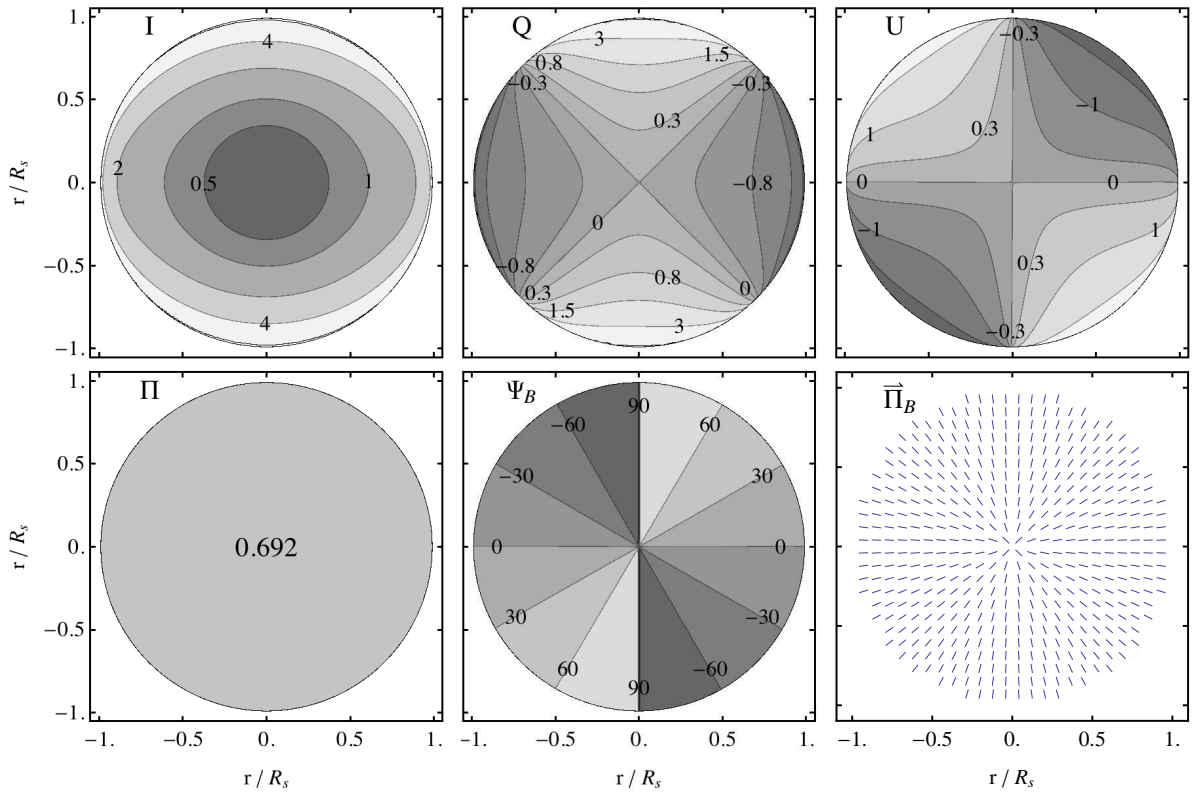}
\caption{Case of radial MF with magnitude $\propto\sin\tht$, and quasi-perpendicular distribution of particles, as seen at an aspect angle of $60^\circ$.}
\label{fig:radialBLang60}
\end{figure*}
\section{Polar-caps vs Barrel-like models}

So far we have considered the simplest possible case, namely that in which the shocked ambient medium is confined to a shell of constant thickness, the MF into that shell is just the result of the ambient field compression, and the normalization of the particles energy spectrum is the same everywhere.

Let us release now the last assumption and, following \citet{Fulbright-Reynolds-1990}, let us parametrize the acceleration efficiency with the ``obliquity angle'', namely the angle between the shock normal and the (ordered component of the) MF.
Consistently with what typically assumed in the literature, we will consider here the angle between the shock velocity and the post-shock MF ($\thtntwo$), which is linked to the polar angle by the relations
\begin{eqnarray}
\label{eq:sinthtntwo}
\sinthtntwo&=&\frac{\comprratio\sintht}{\sqrt{\comprratio^2\sinsqtht+\cossqtht}};	\\
\label{eq:costhtntwo}
\costhtntwo&=&\frac{\costht}{\sqrt{\comprratio^2\sinsqtht+\cossqtht}}.
\end{eqnarray}
Incidentally, the polar angle $\tht$ coincides with the angle between the shock velocity and the ambient MF, $\thtnone$.

Let us also define, as ``quasi-parallel'' and ``quasi-perpendicular'' cases, those in which the normalization of the particle energy distribution is respectively proportional to $\cossqthtntwo$ (therefore enhancing the emissivity at obliquity angles closer to $0^\circ$) or to $\sinsqthtntwo$ (which privilege angles closer to $90^\circ$).
In the former case the emission will be then enhanced near the polar regions (``polar-cap'' morphology), while in the latter one around the equatorial regions (``barrel-like'' morphology).
Critical comparisons of these two morphologies are present in the literature, when discussing the nature of bilateral SNRs in general \cite[e.g.\ ][]{Orlando-etal-2007, Fulbright-Reynolds-1990}, or more specifically the case of SN~1006 \cite[(e.g.\ ][]{Schneiter-etal-2015, Bocchino-etal-2011, Schneiter-etal-2010}.

In fact, the quasi-perpendicular prescription leads to maps very similar to those of the isotropic case: in fact Eq.~\ref{eq:sinthtntwo} gives values very close to unity, except in narrow regions around the poles.
The opposite is for Eq.~\ref{eq:costhtntwo}, which leads to polar-cap enhanced emission, and for which one can easily appreciate the differences  (see for instance Fig.~\ref{fig:meridFOnoFRang90PARA}, for an aspect angle of $90^\circ$, to be compared with Fig.~\ref{fig:meridFOnoFRang90}).

We have already mentioned that, for young SNRs, there is an indication that the MF configuration must be mostly radial.
Our main intention here, far from presenting specific and detailed models for individual SNRs, is to show in general how radio polarization data could be effectively used to investigate similar objects; but, for the sake of illustration, we shall consider here as an archetypal case the remnant of SN~1006, which has been extensively studied, and of which very detailed radio emission and polarization maps are now available.

Polarization mapping of this source \citep[see e.g.][]{Reynolds-Gilmore-1993} show that the direction of the projected MF is approximately radial along the two brighter limbs (in the NE and SW sectors).
This pattern of polarization is clearly inconsistent with a quasi-perpendicular model (if the MF stays ordered and the internal FR is negligible), as shown for instance by \citet{Schneiter-etal-2015}, and may be easily inferred from our Fig.~\ref{fig:meridFOnoFRang90}.
Instead, an ambiguity may remain between a quasi-parallel case with ordered MF, and either a quasi-parallel or a quasi-perpendicular case, but with the formation of radial MF patterns (possibly due to the onset of instabilities), especially in the brighter regions.

More recently, \citet{Reynoso-etal-2013} have shown that the polarization along the two fainter edges of SN~1006 (in the SE and NW sectors) is consistent with a tangential MF, and that while the average fractional polarization of the two bright limbs is about $17\%$, it typically increases along the weaker limbs: for instance, in the SE sector it reaches a value of about $60\pm20\%$ (this last value being consistent with the theoretical limit of $\simeq71\%$, for a radio spectral index $\simeq0.6$).
\citet{Reynoso-etal-2013} interpret these results in the framework of the quasi-parallel case, and in addition take the lower polarization fraction in the two brighter limbs as the signature of efficient particle acceleration and generation of magnetic turbulence in those regions.

However, in Sect.~2.3 we have shown that, in the case of strong turbulence with an isotropic random MF in the upstream, the shock compression would imply a strong anisotropy in the immediate downstream turbulence, leading anyway to large polarization fractions ($\simeq60\%$ for typical radio spectral indices, see Fig.~\ref{fig:PolarizedShock}), with a polarization consistent with a preferential MF direction tangent to the shock front.
Therefore, except for very fine tuned conditions, the observation of typically radial patterns for the projected MF seems to require the onset of Rayleigh-Taylor-like instabilities, and for this reason we are going here to examine the effects of a preferentially radial MF, in the framework of a barrel-like model.

For the sake of simplicity, we shall consider here a monopole-like configuration for the ordered MF.
The radially oriented MF may actually have alternate orientations in different places; but this neither affects the magnitude of the synchrotron emissivity, nor its polarization.
Similarly to the standard quasi-perpendicular model, we shall use also here the $B\propto\sin\tht$ prescription; but, differently from before, here the only justification is that in this way the azimuthal dependence of the radio emission in SN~1006 can be roughly reproduced.
For the normalization of the particle energy distribution we will use again the $\sinsqthtntwo$ prescription, as in the standard quasi-perpendicular case.

In Fig.~\ref{fig:radialBLang60}, we show respectively the Stokes parameters and the polarization in the absence of FR, with an aspect angle of $60^\circ$ \citep[see][for estimates of the aspect angle for SN~1006]{Petruk-etal-2009}.
On the other hand, if as mentioned above the radial MF reverses its orientation many times on smaller scales, one should not be able to detect any net effect of this radial pattern on the FR; while only the FR originating from the meridional structure would not average to zero.
A strategy to tell apart the correct model could then involve an analysis of the pattern of the internal RM, in order to identify the symmetry axis of the underlying meridional MF.
In fact, the pattern of the RM, independently of the distribution of the emitting particles, would allow one to identify the symmetry axis of the MF meridional component (see e.g.\ Fig.~\ref{fig:meridRMFRang60} or \ref{fig:meridRMFRang90} for cases with large aspect angles).
Therefore, if an overall gradient of the RM is detected, it may indicate the direction of the axis of symmetry of the SNR, and therefore if the radio emission from this source shows a bilateral structure it would be interesting to check whether the two directions are nearly parallel or nearly orthogonal.
Of course, this kind of analysis requires the absence of gradients in the foreground FR (to be checked observing nearby polarized background sources) as well as a not too distorted structure for the SNR, therefore limiting the numbers of objects on which such kind of analysis could be actually performed.

Finally, let us discuss the implications of the argument presented by \citet{Rothenflug-etal-2004} to disprove a barrel-like emission structure: stated in a qualitative way, they noted that in a symmetric barrel-like structure one can quantify a lower limit to the ratio of the emission near the projected center of the SNR and from the projected limbs.
More formally, they have first considered a homogeneously emitting annulus of radius $r$, and defined ${\cal R}_{\lmb}$ \citep[here, following the notation in][the symbol $\lmb$ is an angle]{Rothenflug-etal-2004} as the ratio between the flux of the regions whose projected distance from the symmetry axis is less than $r\sin\lmb$, and those with projected distance larger than $r\sin\lmb$: in this case, by purely geometrical arguments, this ratio is found to be equal to $\pi/(2\lmb)-1$; in particular, the authors use $\lmb=\pi/3$, for which the ratio evaluates $0.5$.
Their last point is that a generic distribution of the emission can be imagined as the sum of more annuli, with different radii, in which case using that criterion for the largest radius gives a lower limit to the global ratio ${\cal R}_{\pi/3}>0.5$).

\begin{figure}
\centering
\includegraphics[angle=0,width=8.0truecm]{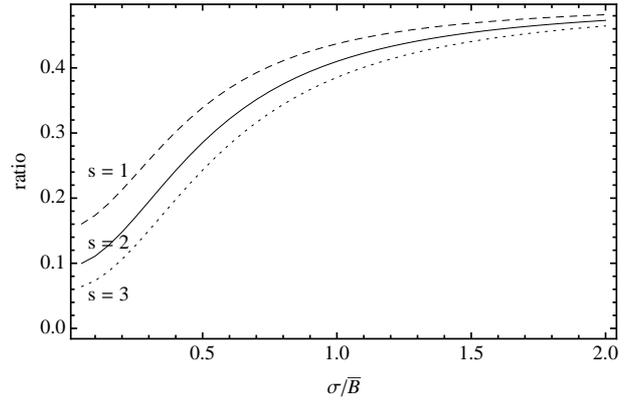}
\caption{The \citet{Rothenflug-etal-2004} criterion, generalized to the case of a radially oriented ordered MF plus a random MF component.}
\label{fig:RothenflugRev}
\end{figure}

\citet{Rothenflug-etal-2004}, in the derivation of their criterion, used the assumption of ``isotropic radiation'' (cf. their Sect. A.1).
This assumption, while apparently reasonable, in fact requires the MF to be completely random and, among others, it would be inconsistent with the presence of polarized emission, as actually observed. On the other hand, if we release this assumption the situation can change considerably.
In particular, Fig.~\ref{fig:RothenflugRev} shows the case of an annulus seen edge-on (now the orientation matters), for different values of the ratio between the random and ordered (radial) MF, for some choices of the spectral index.
It is apparent that, while the asymptotic limit for large $\sg/\Bav$ values is $0.5$, independently of the slope $s$, lower values are obtained for non negligible values of the ordered component.

More insight can be obtained by computing the equatorial radial profile for the model shown in Fig.~\ref{fig:radialBLang60} (with an aspect angle of $60^\circ$), for different values of $\sg/\Bav$.
It can be seen that, while the  emission near the projected center is appreciable for large $\sg/\Bav$ values (mostly random MF), in the limit of a completely ordered radial MF ($\sg/\Bav=0$) it goes to vanish.
In the lower panel of the same figure we show the profile of the fractional polarization.
The fact that in the bright limbs of SN~1006 a polarization fraction of $\sim20\%$ is measured \citep[][]{Reynolds-Gilmore-1993, Reynoso-etal-2013}, this poses an upper limit of $\simeq1$ to $\sg/\Bav$.
Therefore, the actual profile should be somehow dimmer near the center than what estimated for a purely random MF.

In order to allow a closer comparison with the radio profiles shown in \citet{Rothenflug-etal-2004}, in Fig.~\ref{fig:SummedCut} we also present the profiles of the flux, integrated along the direction orthogonal to the projected equator: in this case the differences between models with different $\sg/\Bav$ values are less evident, but still existing.
In particular, one may see from Fig.~\ref{fig:SummedCut} that, in the fully ordered MF case, the integrated $I$ value on the symmetry axis is just about 40\% of the \citet{Rothenflug-etal-2004} case (i.e. $\sg/\Bav=\infty$).
Therefore, the criterion introduced in that paper cannot be intended by itself as a mathematical proof against a barrel-like geometry. In fact, in the case of a radially-oriented ordered component of the MF, it would be oriented almost along the projection plane near the limbs and toward the observer near the center, and this may result in smaller values of the ratio of emission between center and limbs.

\begin{figure}
\centering
\includegraphics[angle=0,width=8.0truecm]{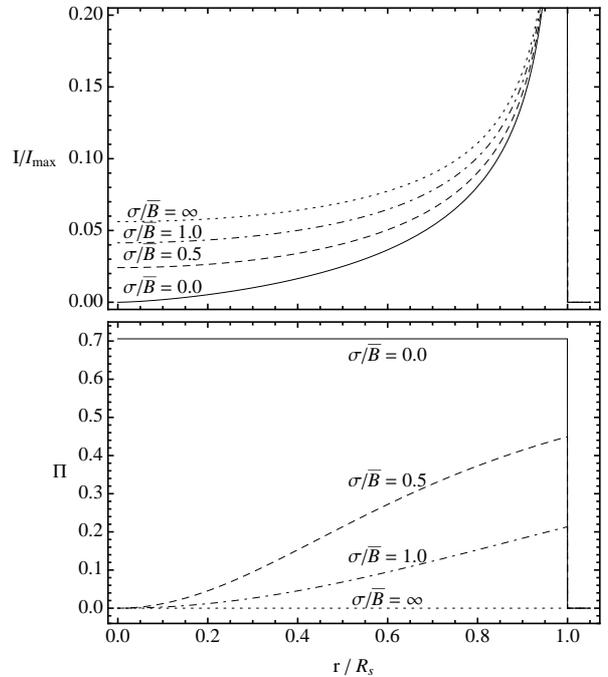}
\caption{Profiles for a model with radial MF with magnitude $\propto\sin\tht$, quasi-perpendicular distribution of particles  (barrel-like model), and different values of $\sg/\Bav$, as seen at an aspect angle of $60^\circ$. Upper panel: normalized $I$ flux profiles, for an equatorial cut. Lower panel: profiles of the polarization fraction, for the same cases.}
\label{fig:Cut}
\end{figure}
\begin{figure}
\centering
\includegraphics[angle=0,width=8.0truecm]{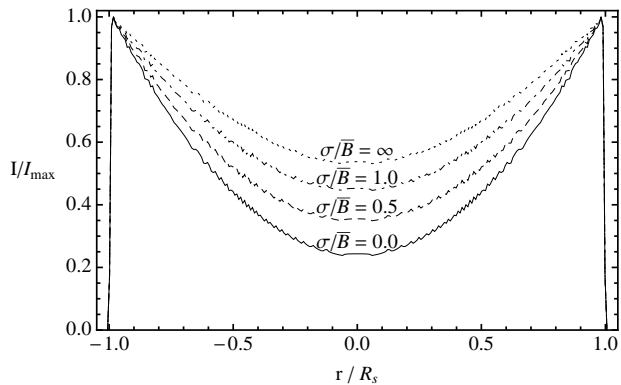}
\caption{Normalized $I$ flux profiles, integrated along the directions perpendicular to the equator, for the same models as in Fig.~\ref{fig:Cut}.}
\label{fig:SummedCut}
\end{figure}

\section{Summary}

In this paper we have presented a generalization of the classical treatment of the synchrotron radiation, valid in the case in which the MF has a random component.
The resulting formulae, while nicely resembling those in the classical case, are now valid for any relative level between ordered and random component, and clearly show that the emission in the case a partially random case cannot be simply modelled as the sum of the emission in a fully ordered MF and that in a completely random MF.

We have then used these formulae to model the synchrotron emission from shell-type SNRs.
The main limitation of the present treatment is its applicability only to the case in which the particle energy distribution is a power law, and for this reason we have focussed our attention on the radio emission in SNRs.
For the future we plan to extend this treatment also to particle energy distributions with an upper cutoff, and therefore to allow also an analysis of the non-thermal SNR emission in the X-ray energy range.
Another limitation of this treatment is that the fluctuations considered are on scales smaller that the instrumental resolution; while for fluctuations on larger scales the only possible approach is through multidimensional, high resolution, fully numerical simulations. 

Our goal here is instead to allow a numerically much lighter treatment than full simulations, while still keeping enough detail to allow investigations of several effects.
With this aim, in order to model the emission from shell-type SNRs we have introduced a thin-shell model.
Our discussion has involved various issues, like the dependence of the typical polarized fractions on the level of MF fluctuations, and the pattern of the polarization as well as of the RM on the geometry of the ordered MF component.
We have also introduced the idea that in the presence of internal FR the measured RM maps contain an information different from that of the classical case, when FR is just due to a foreground medium.
In addition, we have also discussed further effects, on the observed direction of the polarization as well as on its level, which should be more evident at longer radio wavelengths and which, if suitably tested in the future, could provide additional information on the MF structure. 

While this paper has been mostly devoted to cases with a meridional geometry of the ordered MF, as one would expect in the case of a laminar expansion into an ambient medium with a pre-existing homogeneous MF, having in mind the case of young SNRs, and SN~1006 in particular, we have also discussed the case of a radially oriented MF.
We have shown that the radial profiles of the emission may effectively depend also on the level of MF fluctuations, so that in the case of a low level of fluctuations also a barrel-like geometry of the emissivity could be consistent with lower surface brightness near the central regions of the projected image.

\section*{Acknowledgements}

We thank the referee, Dr. P.F. V\'elazquez, for his careful reading and a number of suggestions and corrections that helped us to improve the paper considerably.
This work is partially funded by the PRIN INAF 2010 and by the CNRS-INAF PICS 2012.






\bsp	
\label{lastpage}
\end{document}